\documentclass[10pt, a4paper]{article}

\usepackage{graphicx}
\usepackage{tikz}
\usetikzlibrary{positioning, fit, backgrounds, arrows.meta, calc}

\tikzset{
    inputbox/.style={
        rectangle,
        rounded corners,
        draw=black!70,
        fill=gray!10,
        minimum width=2.2cm,
        minimum height=0.75cm,
        align=center,
        font=\small
    },
    mlpbox/.style={
        rectangle,
        rounded corners,
        draw=blue!70,
        fill=blue!5,
        minimum width=2.6cm,
        minimum height=0.9cm,
        align=center,
        font=\small
    },
    latentbox/.style={
        rectangle,
        rounded corners,
        draw=red!70,
        fill=red!5,
        minimum width=1.8cm,
        minimum height=0.75cm,
        align=center,
        font=\small
    },
    headbox/.style={
        rectangle,
        rounded corners,
        draw=green!60!black,
        fill=green!5,
        minimum width=1.7cm,
        minimum height=0.55cm,
        align=center,
        font=\small
    },
    physicsbox/.style={
        rectangle,
        rounded corners,
        draw=purple!70,
        fill=purple!5,
        minimum width=2.2cm,
        minimum height=0.7cm,
        align=center,
        font=\small
    },
    pdebox/.style={
        rectangle,
        rounded corners,
        draw=red!80,
        fill=red!5,
        minimum width=7.8cm,
        minimum height=1.2cm,
        align=center,
        font=\small
    },
    opcircle/.style={
        circle,
        draw=black!70,
        fill=white,
        minimum size=0.65cm,
        align=center,
        font=\small
    },
    arrow/.style={
        -{Latex[length=2mm]},
        thick
    },
    dasharrow/.style={
        -{Latex[length=2mm]},
        thick,
        dashed
    }
}

\usepackage[margin=1in]{geometry} 
\usepackage{authblk} % 用于处理多作者和多机构排版
\usepackage{authblk}
\usepackage{tikz}
\usetikzlibrary{fit, backgrounds, positioning}

\usepackage[fleqn]{amsmath} % fleqn 让公式左对齐，保留了你原来的设置
\usepackage{amsthm}
\usepackage{amssymb}

\usepackage{graphicx}
\usepackage{subcaption}
\usepackage{float}
\usepackage[numbers, sort&compress]{natbib}
\usepackage{booktabs}
\usepackage[table]{xcolor}

\usepackage{comment}
\usepackage{datetime2}
\usepackage{tikz}
\usetikzlibrary{shapes.geometric, arrows.meta, positioning, fit, backgrounds, calc}
\usepackage[colorlinks=true, allcolors=blue]{hyperref} % 加上了颜色，方便你在草稿期阅读
\usepackage[nameinlink,noabbrev]{cleveref}

\begin{document}

\title{Yield Curve Dynamics Using Variational Autoencoders Under No-arbitrage}

% ==========================================
% 作者与机构 (Authors & Affiliations)
% 使用 authblk 宏包的标准写法，日后无缝切换各大期刊模板
% ==========================================
\author[1]{Fusheng Luo\thanks{Work completed while the author was at Johns Hopkins University. The author is currently an Independent Researcher.}}
\author[2]{H\'elyette Geman}

% 建议在这里填写真实的机构信息，例如：
\affil[1]{Department of Applied Mathematics and Statistics, Johns Hopkins University, USA}
\affil[2]{Department of Applied Mathematics and Statistics, Johns Hopkins University, USA}

% 如果不想让机构字号太大，可以加上这句调整格式：
% \renewcommand\Affilfont{\itshape\small}

\date{\today}

\maketitle

% ==========================================
% 摘要 (Abstract)
% 标准 LaTeX 中，摘要必须放在 \maketitle 之后，并使用 abstract 环境
% ==========================================
\begin{abstract}
This paper introduces a physics-informed generative framework that resolves the fundamental conflict between the statistical flexibility of deep learning and the rigorous theoretical constraints of fixed-income modeling. We demonstrate that standard generative models and unconstrained statistical extrapolations suffer from "manifold collapse" and severe arbitrage violations when forecasting term structures across diverse macroeconomic regimes. To overcome this, we propose a two-stage architecture. First, a Student-t Conditional Variational Autoencoder with Dynamic Level Injection ($\text{CVAE}_{\text{sT}} + \text{LS}$) extracts a robust, heavy-tailed term structure manifold, effectively decoupling macroeconomic shape dynamics from absolute base rates. Second, the latent dynamic evolution is governed by a continuous-time Neural Stochastic Differential Equation (SDE) strictly penalized by a No-Arbitrage Partial Differential Equation (PDE). Empirical results across multiple sovereign currencies (USD, GBP, JPY) confirm that our synergistic approach drastically reduces out-of-sample forecasting errors—achieving an exceptional 6.58 bps Mean Tenor RMSE—and successfully overcomes the massive parallel drift and zero-lower-bound violations exhibited by the classical HJM model in extreme environments. Furthermore, through phase space vector field analysis, we demonstrate the model's superior capability in unsupervised macroeconomic regime detection and high-quality continuous-time scenario generation. Ultimately, this research provides a highly scalable, mathematically sound evolutionary engine for term structure modeling.
\end{abstract}

% ==========================================
% Keywords
% ==========================================
\vspace{1em}

\vspace{2em}

%% ================================
%% ================================

\section{Introduction}
\label{sec:introduction}

Modeling the dynamic evolution of the yield curve is a cornerstone of modern quantitative finance, essential for macroeconomic policy formulation, portfolio risk management, and the pricing of interest rate derivatives. Over the past decade, and particularly culminating in the post-2022 global monetary tightening cycle, sovereign fixed-income markets have exhibited unprecedented structural volatility. The synchronized departure from the Zero-Lower-Bound (ZLB) in Western economies (e.g., USD, GBP) and the persistent, highly constrained Yield Curve Control (YCC) regimes in Asia (e.g., JPY) have generated extreme, heavy-tailed market shocks. These abrupt regime shifts have critically exposed the theoretical limitations of classical term structure models and modern statistical extrapolation techniques alike.

Historically, quantitative term structure modeling has been bifurcated into two mutually exclusive paradigms. The first paradigm relies on rigorous continuous-time financial mathematics, epitomized by short-rate models (Vasicek \cite{vasicek1977equilibrium}, Cox-Ingersoll-Ross \cite{cox1985theory}) and the Heath-Jarrow-Morton (HJM) framework \cite{heath1992bond}. While these models perfectly enforce the absence of arbitrage by definition, they depend heavily on rigid, low-dimensional parametric assumptions (typically Gaussian Brownian motions). Consequently, they inherently fail to absorb heavy-tailed macroeconomic shocks, frequently resulting in massive parallel drift out-of-sample or catastrophic failures near the zero-lower-bound. 

The second paradigm, driven by the recent explosion of deep learning in quantitative finance, attempts to model the yield curve purely via non-linear state-space models and generative architectures, such as Variational Autoencoders (VAEs) \cite{bergeron2021variationalautoencodershandsoffapproach}. While deep generative models exhibit extraordinary statistical flexibility in capturing historical shape variations (Level, Slope, and Curvature) without relying on explicit analytical forms like the Nelson-Siegel family [\cite{dybvig1996long}, \cite{nelson1987parsimonious}, \cite{christensen2011affine}, \cite{svensson1994estimating}], they are fundamentally ``blind'' to the laws of physics that govern financial markets. Unconstrained by risk-neutral pricing measures, these black-box models frequently suffer from "manifold collapse" during market stress and routinely generate cross-sectional predictions that severely violate fundamental no-arbitrage conditions. This inability to maintain mathematical consistency across maturities is known as the Filipovi\'c consistency problem \cite{filipovic2001consistency}.

To bridge this fundamental gap, this paper proposes a physics-informed generative framework that seamlessly fuses the statistical flexibility of deep learning with the rigorous gravity of continuous-time fixed-income mathematics. We propose a synergistic, two-stage architecture. In Stage A, we design a Student-t Conditional Variational Autoencoder equipped with a Dynamic Level Injection mechanism ($CVAE_{sT} + LS$). This module extracts a robust, low-dimensional continuous manifold by effectively decoupling absolute macroeconomic base rates from relative curve shape dynamics. The Student-t likelihood explicitly accommodates heavy-tailed market anomalies, preventing the topological collapse of the latent space during severe monetary policy shocks. 

In Stage B, rather than relying on unconstrained autoregressive forecasting, the temporal evolution of the latent manifold is governed by a continuous-time Neural Stochastic Differential Equation (SDE). Crucially, this SDE is strictly penalized by a No-Arbitrage Partial Differential Equation (PDE) evaluated directly via exact Automatic Differentiation (Autograd). By enforcing the risk-neutral martingale condition on the decoded discount factors, the Neural SDE computationally solves the Filipovi\'c tangency requirement, acting as a non-linear ``physical gravity'' that forces forecasted curves back into theoretically sound, arbitrage-free states.

Empirical evaluations across multiple sovereign regimes confirm that our physics-informed framework drastically reduces out-of-sample forecasting errors. The proposed architecture achieves an exceptional 6.58 bps Mean Tenor RMSE, significantly outperforming classical multi-factor HJM models and unconstrained statistical baselines. Furthermore, the model naturally learns the highly asymmetric boundary-bounce dynamics required for the Japanese Yen (JPY), perfectly respecting the zero-lower-bound without necessitating the rigid analytical constraints of classical models.

In summary, the core contributions of this paper are threefold:
\begin{itemize}
    \item \textbf{Decoupling Macro-levels via LevelScript (LS):} We introduce a deterministic physical anchor prior to the VAE encoding phase, completely eradicating the catastrophic parallel drifts and base-rate memorization issues that plague standard generative term structure models.
    \item \textbf{Heavy-Tailed Manifold Resilience:} We demonstrate that substituting Gaussian assumptions with a Student-t likelihood within the CVAE prevents manifold collapse, allowing the network to gracefully absorb massive, non-linear market shocks.
    \item \textbf{Exact No-Arbitrage SDE via Autograd:} We engineer a continuous-time Neural SDE that is dynamically regularized by a strict No-Arbitrage PDE. By leveraging the exact Hessian-vector products of the deep decoder, we ensure the empirical manifold remains strictly invariant under risk-neutral pricing laws.
\end{itemize}

The remainder of this paper is organized as follows. Section \ref{sec:background} reviews the theoretical background of interest rate modeling and the Filipovi\'c no-arbitrage condition. Section \ref{sec:data} describes the multi-currency OIS swap dataset and the shape-level decomposition. Section \ref{sec:vae_variants} and \ref{sec:methodology} formalize the architecture of the $CVAE_{sT}+LS$ and the No-Arbitrage Neural SDE. Section \ref{sec:benchmarks} outlines the advanced theoretical benchmarks. Section \ref{sec:results} presents the empirical evaluation, ablation studies, and phase space vector field analysis. Section \ref{sec:conclusion} concludes the paper.

\section{Theoretical Background and Related Work}
\label{sec:background}

To contextualize the architectural choices of the proposed framework, we first review the classical evolution of term structure modeling. We then define the fundamental no-arbitrage constraints under the Heath-Jarrow-Morton (HJM) framework \cite{heath1992bond} and articulate the Filipovi\'c consistency problem \cite{filipovic2001consistency}---the exact mathematical bottleneck that our physics-informed neural network is designed to solve.

\subsection{Evolution of Term Structure Models}
The modeling of interest rates has historically diverged into two distinct methodological paths: equilibrium short-rate models and deterministic shape-fitting models. Early continuous-time frameworks, such as those proposed by Vasicek \cite{vasicek1977equilibrium} and Cox-Ingersoll-Ross (CIR) \cite{cox1985theory}, characterize the yield curve dynamically driven by a low-dimensional Markovian state variable (the instantaneous short rate). While tractable, these models often lack the flexibility to perfectly calibrate to the initial term structure.

Conversely, practitioners and central banks have extensively relied on parsimonious deterministic frameworks, most notably the Nelson-Siegel (NS) \cite{nelson1987parsimonious} and Nelson-Siegel-Svensson (NSS) \cite{svensson1994estimating} models. These models map the cross-section of yields onto a set of interpretable macroscopic factors: level, slope, and curvature. Despite their empirical success and the strong support from principal component analysis (PCA) \cite{litterman1991volatility}, static NS/NSS models are inherently vulnerable when deployed in a dynamic forecasting setting. As demonstrated by Christensen et al. \cite{christensen2011affine}, dynamic implementations of standard Svensson factor-loadings do not automatically guarantee inter-temporal no-arbitrage restrictions.

To unify calibration flexibility with strict pricing theory, Heath, Jarrow, and Morton (HJM) \cite{heath1992bond} introduced an infinite-dimensional framework that models the evolution of the entire instantaneous forward rate curve $f(t,T)$. Assuming frictionless markets and the existence of an Equivalent Martingale Measure (EMM) $\mathbb{Q}$, the fundamental HJM restriction dictates that the risk-neutral drift of the forward curve is entirely determined by its volatility structure.

\subsection{Deep Generative Models in Fixed Income}
The rigid parametric boundaries of classical models have recently been challenged by the application of deep generative architectures. Variational Autoencoders (VAEs) \cite{kingma2013auto} have proven highly effective at learning low-dimensional, non-linear manifolds of yield curve shapes directly from historical swap rates or zero-coupon yields. Recent literature, including the works of Sokol \cite{sokol2022autoencoder} and Lyashenko et al. \cite{lyashenko2024autoencoder}, illustrates that VAEs can effectively compress high-dimensional forward curves into a compact set of latent factors while discarding idiosyncratic noise.

However, a critical limitation persists: when unconstrained standard VAEs (typically relying on Gaussian priors) encounter heavily biased datasets or unprecedented macroeconomic regimes, they frequently suffer from "prior holes" or manifold collapse \cite{saha2025matching}. Manifold collapse denotes the fact that in order to fit into the extreme values in the abnormal financial period, the model will lose its general 
ability to produce normal and reasonable yield rates at stable financial periods. Furthermore, while these models capture static cross-sectional dependencies elegantly, imposing continuous-time, arbitrage-free temporal dynamics onto their ``black-box'' latent spaces remains a largely unresolved challenge in the literature.

\subsection{The Filipovi\'c Consistency Problem and Neural Manifolds}
The empirical success of our two-stage architecture is mathematically grounded in resolving the consistency theory for HJM models, formally proved by Filipovi\'c \cite{filipovic2001consistency}. 

To enhance the HJM framework for practical simulation, Musiela \cite{musiela1993different} parameterized the curve using time-to-maturity coordinates, $x = T - t$. The forward curve $r_t(x)$ evolves as a Stochastic Partial Differential Equation (SPDE):
\begin{equation}
    dr_t(x) = \left( \frac{\partial}{\partial x}r_t(x) + \sum_{i=1}^n \sigma_i(t,x)\int_0^x \sigma_i(t,u)du \right)dt + \sum_{i=1}^n \sigma_i(t,x)dW_{i,t}^{\mathbb{Q}}
\end{equation}

Filipovi\'c investigated under what conditions this infinite-dimensional SPDE admits a finite-dimensional realization. He proved that a finite-dimensional parameterized manifold $\mathcal{M} = \{G(z) | z \in \mathcal{Z} \subset \mathbb{R}^d\}$ is invariant (or consistent) with the HJM dynamics \textit{if and only if} the drift and volatility vector fields of the SPDE are strictly tangent to $\mathcal{M}$ at every point. Crucially, Filipovi\'c proved that the widely used Nelson-Siegel family does \textit{not} form a consistent manifold, meaning that an NS curve will almost surely evolve into an economically implausible, non-NS curve under arbitrage-free continuous dynamics. We discuss more regarding this topic at section \ref{sec:appendix_theory}.

Our architecture provides a data-driven computational solution to this exact problem. Rather than guessing analytical families like exponential-polynomials, Stage A uses a frozen, heavy-tailed decoder $D^{(P)}(z, \tau)$ to define a highly expressive, $d$-dimensional continuous neural manifold:
\begin{equation}
    \mathcal{M}_{Neural} = \{D^{(P)}(z, \cdot) | z \in \mathbb{R}^d\}
\end{equation}
Because this manifold is learned directly from empirical OIS data, it naturally spans the true topological space of observed macroeconomic shapes. 

In Stage B, for $\mathcal{M}_{Neural}$ to act as a consistent HJM manifold under Filipovi\'c's definition, the stochastic evolution of the discounted bond prices generated by $D^{(P)}$ must be a local martingale under $\mathbb{Q}$. As we will detail in Section \ref{sec:methodology}, our No-Arbitrage PDE loss ($\mathcal{L}_{arb}$) directly penalizes deviations from Filipovi\'c's tangency requirement. By enforcing this penalty, the Neural SDE ensures that the local dynamics of the bond price lie perfectly within the tangent bundle spanned by the deep decoder's Jacobian, thereby achieving both empirical accuracy and theoretical invariance.

%% ================== Section: Dataset ==================
\section{Dataset and Term Structure Geometry}
\label{sec:data}

\subsection{Data Preprocessing and Stable-Coverage Truncation}
We analyze a comprehensive panel of daily Overnight Index Swap (OIS) curves across eight sovereign regimes. To capture the full spectrum of term structure geometry while retaining cross-market comparability, we define a canonical maturity grid of $|\mathcal{T}| = 12$ tenors:
\begin{equation}
    \mathcal{T} = \{1\text{M}, 2\text{M}, 3\text{M}, 6\text{M}, 1\text{Y}, 2\text{Y}, 5\text{Y}, 7\text{Y}, 10\text{Y}, 15\text{Y}, 20\text{Y}, 30\text{Y}\}
\end{equation}
This grid captures the short-end slope, intermediate hump/curvature, and long-end behavior with sufficient granularity for representation learning in low latent dimensions ($2$--$3$). The choice also avoids non-standard maturities that are consistently absent in some markets. Details of how we deal with the null values can be found in our Appendix \ref{sec:appendix}. 

Figure \ref{fig:daily_multicurrency} visualizes cross-sectional term-structure shapes by plotting $20$ randomly sampled daily curves for each currency on the common tenor grid. The overlaid lines illustrate the diversity of curve regimes (level and slope) within each market, including both upward- and downward-sloping configurations. This diagnostic is used to validate data coverage and to assess whether the selected tenor set captures meaningful variation in curve shape prior to the model training. 

\subsection{The LevelScript Mechanism and Shape-Level Decomposition}
Figure \ref{fig:1y_multicurrency} plots the time series of the $1$-year swap rate (used as the LevelScript (LS) factor) for each currency over its available post-truncation sample. The panels highlight distinct monetary-policy regimes and structural level shifts across markets—e.g., prolonged low/negative-rate environments (JPY, CHF) and the synchronized global tightening episode around $2022$--$2023$. 

Because term-structure levels differ substantially across currencies and regimes, feeding raw rates would cause a latent model to allocate capacity primarily to explaining cross-market level differences, rather than learning shape dynamics. To ensure interpretability and stable learning, we construct a \emph{level script mechanism}. Let the level factor be defined as the $1$-year rate:
\begin{equation}
l_t^{(c)} = r_t^{(c)}(\text{1Y})
\end{equation}

We define the shape vector as the curve expressed relative to this anchor:
\begin{equation}
s_t^{(c)} = r_t^{(c)}(\tau) - l_t^{(c)}, \quad \text{for } \tau \in \mathcal{T}
\end{equation}

By construction, $s_t^{(c)}(\text{1Y})$ and the remaining dimensions represent relative slope and curvature features independent of level. To avoid undue influence from outliers or regime shifts, we apply robust scaling separately to the shape coordinates and the level coordinate. For a generic variable $y$, robust scaling is performed using the median and interquartile range (IQR):
\begin{equation}
\tilde{y} = \frac{y - \text{median}(y)}{\text{IQR}(y)}
\end{equation}
This normalization is computed on the pooled (cross-currency) post-truncation dataset, producing dimension-wise comparable inputs while preserving the statistical structure of the panel. The final model input concatenates a scaled shape vector and a scaled level scalar $x_t^{(c)} = [\tilde{s}_t^{(c)}, \tilde{l}_t^{(c)}]$ yielding a fixed-dimensional input suitable for a single VAE.

Figure \ref{fig:pca_multicurrency} visualizes the dominant modes of variation in the shape component of the yield-curve panel. We project all curve observations onto the first two principal components (PC$1$--PC$2$) to assess the intrinsic dimensionality of shape dynamics. The reported explained-variance ratios indicate that PC$1$ and PC$2$ capture the vast majority of shape variation (about 61.7\% and 35.6\%, respectively), supporting the statements of Andreasen \cite{Andreasen2023Decoding}, Sokol \cite{sokol2022autoencoder} and Lyashenko \cite{lyashenko2024autoencoder}: the use of a low-dimensional latent representation ($2$--$3$ factors) is highly optimal for the subsequent VAE.

\section{Variational Autoencoders And Their Variants}
\label{sec:vae_variants}

Variational Autoencoders (VAEs), introduced by Kingma \cite{kingma2013auto}, are generative models that learn a low-dimensional latent space representation of complex data. Unlike traditional autoencoders, which map input to a single fixed point, VAEs map input to a probability distribution. The goal of a VAE is to maximize the likelihood of the observed data $x$ by learning a set of latent variables $z$. Because the true posterior $p(z|x)$ is usually intractable, VAEs use an inference network (encoder) to approximate it with $q_{\phi}(z|x)$. The objective function, known as the Evidence Lower Bound (ELBO), is derived as:
\begin{equation}
    \mathcal{L}(\theta, \phi; x) = \mathbb{E}_{q_{\phi}(z|x)}[\log p_{\theta}(x|z)] - D_{KL}(q_{\phi}(z|x) || p(z)) \label{equ:VAE_loss}
\end{equation}
\begin{equation}
    D_{KL}(q_{\phi}(z|x) || p(z)) = \mathbb{E}_{z \sim q_{\phi}} \left[ \log \frac{q_{\phi}(z|x)}{p(z)} \right]= \mathbb{E}_{z \sim q_{\phi}} [ \log q_{\phi}(z|x) - \log p(z) ] \label{eq:KL_loss}
\end{equation}

With the prior and posterior defined as:
\begin{enumerate}
    \item \textbf{Prior:} 
    \begin{equation}
    p(z) = \mathcal{N}(z; \mathbf{0}, \mathbf{I}), \quad \log p(z) = -\frac{J}{2}\log(2\pi) - \frac{1}{2}\sum_{j=1}^{J} z_j^2
    \end{equation}
    \item \textbf{Approximate Posterior:} $q_{\phi}(z|x) = \mathcal{N}(z; \boldsymbol{\mu}, \boldsymbol{\sigma}^2\mathbf{I})$
    \begin{equation}
        \log q_{\phi}(z|x) = -\frac{J}{2}\log(2\pi) - \frac{1}{2}\sum_{j=1}^{J} \log(\sigma_j^2) - \frac{1}{2}\sum_{j=1}^{J} \frac{(z_j - \mu_j)^2}{\sigma_j^2}
        \label{eq:KL_normal}
    \end{equation}
\end{enumerate}

The Kullback-Leibler (KL) divergence term in Equation (\ref{equ:VAE_loss}) and (\ref{eq:KL_loss}) serves as a critical topological constraint. By penalizing deviations of the approximate posterior from the isotropic Gaussian prior in Equation (\ref{eq:KL_normal}), it enforces a dense and continuous latent space. In the context of yield curve modeling, this regularization ensures that the learned latent factors vary smoothly, preventing the generation of degenerate or economically implausible term structures.

\subsection{The Bottleneck of the Normality Assumption}
However, the academic consensus has increasingly identified this "normality assumption" as a significant bottleneck when modeling financial time series characterized by extreme volatility. The use of a fixed, uninformative Gaussian prior implicitly assumes that the underlying data manifold is Euclidean and that the latent factors follow a unimodal, symmetric distribution. 

As Saha \cite{saha2025matching} and Dutta \cite{dutta2025learning} have pointed out, a significant limitation is the \emph{prior hole problem}. In practice, the Gaussian prior often fails to align with the complex, multimodal distribution of the encoded data, especially in our multi-currency dataset. This creates regions in the latent space that have high probability under the prior but zero density under the aggregated posterior. When sampling from these "holes" during generative simulation, the decoder is forced to interpret latent codes it never encountered during training, resulting in unrealistic shape samples.

\subsection{Architectural Advancements: CVAE, Hyperspherical, and FinQ}
To address non-stationarity, we implement Conditional Variational Autoencoders (CVAEs). Sokol \cite{sokol2022autoencoder} demonstrated the nuanced advantages of CVAEs for multi-currency datasets. By concatenating a Currency-Specific Embedding to both the encoder and decoder inputs, the architecture achieves a decoupling of global latent factors from idiosyncratic market levels. The encoder $q(z|X, C)$ processes the full yield curve alongside a currency-ID vector, ensuring the latent space $z$ captures invariant structural components shared across global markets, while the conditional decoder $p(X|z, C)$ reintroduces currency-specific structural biases. This significantly enhances Out-of-Sample (OOS) robustness (Figure 4).

Other advanced architectures attempt to solve the prior mismatch in different ways. Davidson \cite{davidson2018hyperspherical} proved that a hyperspherical latent space using the von Mises-Fisher (vMF) distribution avoids the "origin gravity" of Gaussian priors. However, the surface area of a hypersphere vanishes as dimensionality increases (the "soap bubble effect"), causing severe numerical issues. Alternatively, Boier \cite{boier2023multiresolution} introduced FinQ-VAEs, a pipeline of cascading VAEs with latent space quantization that snaps latent vectors to market anchor points. While effectively smoothing noise, the selection of anchor points is user-defined rather than automated, and whether calibrating solely to real market data truly eliminates arbitrage in less liquid regimes remains questionable.

\subsection{The Student-t Likelihood Solution}
Given these structural challenges, we adopt the methodology of Takahashi \cite{takahashi2018student}, who identified that standard Gaussian VAEs become hyper-unstable during training on biased datasets. When the neural network estimates a variance ($\sigma^2$) that is almost zero, the Gaussian log-likelihood becomes hypersensitive. Because $\sigma^2$ is in the denominator, even a microscopic error between the true data point ($x$) and the model's predicted mean ($\mu$) returns an extremely large penalty, causing the loss to jump drastically. 

\begin{figure}[H]
  \centering
  \includegraphics[width=0.5\textwidth]{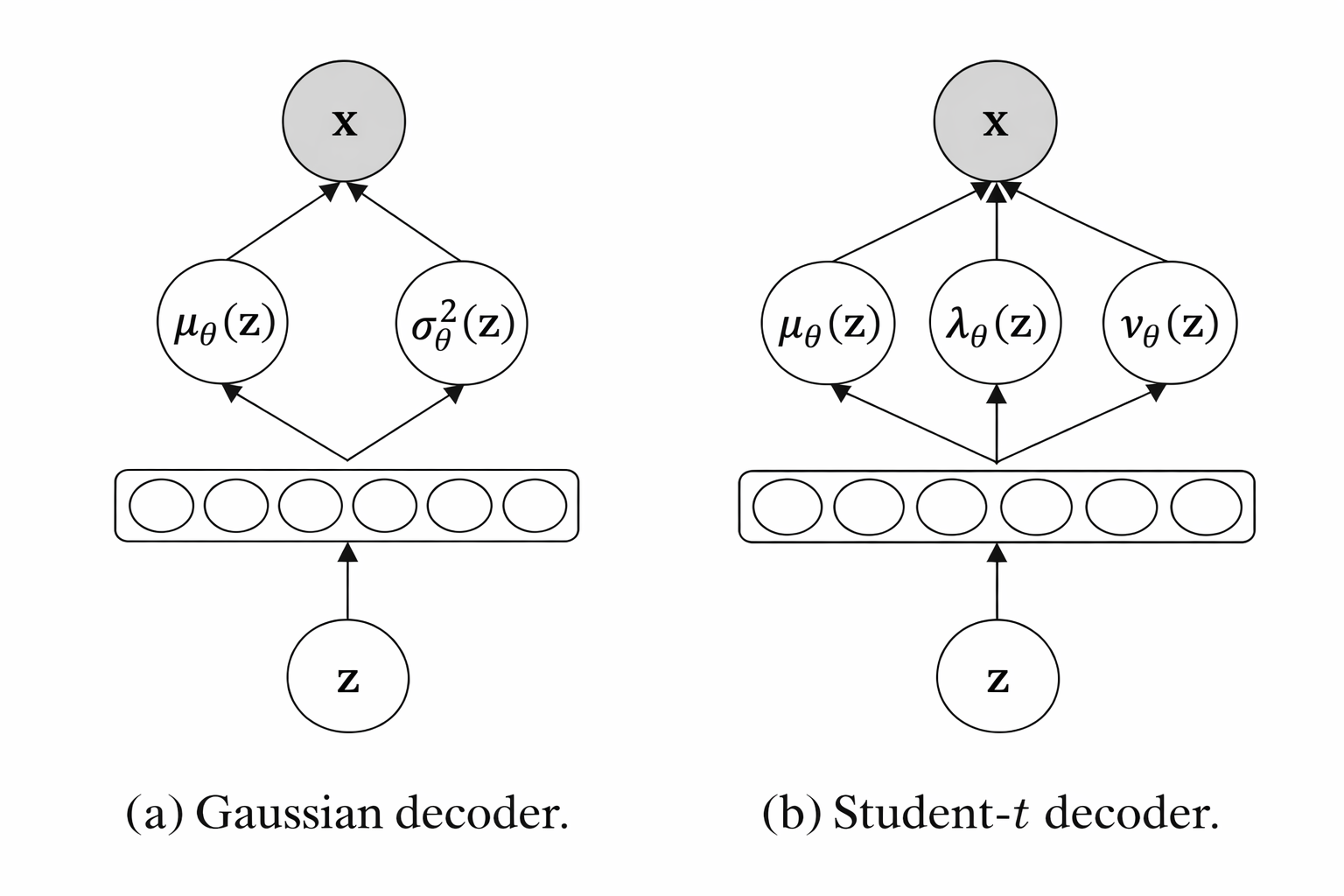}
  \caption{(a) \textbf{Gaussian decoder architecture}: A latent variable $z$ is mapped through a shared hidden representation to predict the mean $\mu_\theta(z)$ and variance $\sigma_\theta(z)$ of the Gaussian likelihood for $x$; (b) \textbf{Student-$t$ decoder architecture}: A latent variable $z$ is mapped through a shared hidden representation to predict the location $\mu_\theta(z)$ scale/precision-related parameter $\lambda_\theta(z)$ and degrees-of-freedom parameter $v_\theta(z)$ of the Student-$t$ likelihood of $x$}
  \label{fig:student_t_decoder}
\end{figure}

To resolve this, we employ a Bayesian marginalization framework. By placing a Gamma prior distribution on the precision parameter and allowing its shape and rate parameters to depend directly on the latent variable $z$, the marginalized loss function follows a Student-$t$ distribution. The explicit reconstruction term (the log-likelihood of the Student-$t$ distribution for a single dimension) is analytically derived as: 
\begin{equation}
    \ln p_\theta(x|z) = \ln \Gamma\left(\frac{\nu_\theta(z)+1}{2}\right) - \ln \Gamma\left(\frac{\nu_\theta(z)}{2}\right) + \frac{1}{2} \ln \left(\frac{\lambda_\theta(z)}{\pi \nu_\theta(z)}\right) - \frac{\nu_\theta(z)+1}{2} \ln \left(1 + \frac{\lambda_\theta(z)(x - \mu_\theta(z))^2}{\nu_\theta(z)}\right)
\end{equation}
where $\Gamma()$ is the gamma function, $\nu_\theta(z)$ is the degrees of freedom of the latent space $z$, the location $\mu_\theta$ represents the expected swap rates, and the scale $\lambda_\theta$ is a proxy for the precision. Figure \ref{fig:student_t_decoder} shows the architectural differences. The characteristics of the Student-$t$ distribution bring heavier tails that perfectly suit our datasets, naturally absorbing the aggressive volatility and outliers inherent across diverse global currency regimes.

\subsection{Model Implementation Details}
\label{sec:model_architecture}
Table \ref{tab:architecture} provides the comprehensive, layer-by-layer architectural blueprint of our two-stage network setup, detailing the tensor dimensions, block compositions, and activation profiles. 

In Stage A, the $CVAE_{sT}$ Encoder ingests a 21-dimensional concatenated vector comprising the curve shape, LevelScript anchor, and currency embedding. It utilizes Linear layers integrated with BatchNorm and Smooth Softplus activations to map the input down to the parameters of the latent distribution ($\mu_z$ and $\log\sigma_z^2 \in \mathbb{R}^3$). The Student-$t$ Decoder takes the 11-dimensional conditional latent vector and maps it back to the three heads governing the curve's predictive distribution: Location ($\mu_\theta$), Scale ($\lambda_\theta$), and Degrees of Freedom ($\nu_\theta$). Critically, to preserve mathematical validity under continuous financial modeling, the DoF head is constrained via a shifted Softplus function to enforce $\nu_\theta > 2$, strictly guaranteeing the existence of a well-defined conditional variance matrix.

Stage B parameters the continuous-time evolution via the Neural SDE ParamNet over the frozen Stage A latent manifold. Operating on the 11-dimensional conditional latent input, the network uses bounded $Tanh$ activations—ensuring smooth, Lipschitz-continuous fields with stable derivatives—to simultaneously output the physical drift $\mu_{\mathbb{P}}$, the diagonal elements of the volatility matrix $\Sigma$, and the market risk premium vector $\lambda$. This highly structured, decoupled parameterization allows for the exact calculation of the risk-neutral drift $\mu_{\mathbb{Q}}$ and enables the analytical evaluation of the No-Arbitrage PDE loss.

\section{Physics-Informed Continuous-Time Dynamics}
\label{sec:methodology}

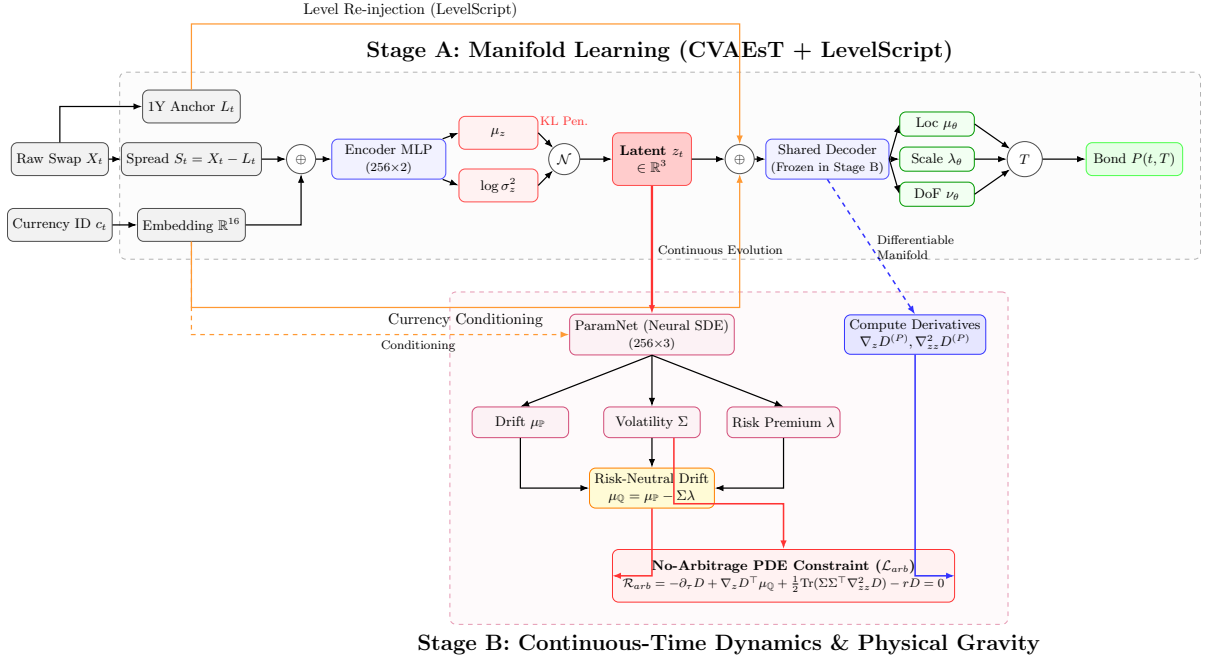
\begin{figure*}[htbp]
    \centering
    \resizebox{\textwidth}{!}{
    \begin{tikzpicture}[node distance=1.2cm]
        % 1. Input & Preprocessing
        \node[inputbox] (raw) at (0, 0) {Raw Swap $X_t$};
        \node[inputbox] (currency) at (0, -1.5) {Currency ID $c_t$};
        
        \node[inputbox] (level) at (3, 1.2) {1Y Anchor $L_t$};
        \node[inputbox] (spread) at (3, 0) {Spread $S_t = X_t - L_t$};
        \node[inputbox] (embedding) at (3, -1.5) {Embedding $\mathbb{R}^{16}$};
        
        \draw[arrow] (raw) |- (level);
        \draw[arrow] (raw) -- (spread);
        \draw[arrow] (currency) -- (embedding);
        
        % 2. Encoder
        \node[opcircle] (concat1) at (5.5, 0) {$\oplus$};
        \node[mlpbox] (encmlp) at (7.5, 0) {Encoder MLP \\ \footnotesize(256$\times$2)};
        \node[latentbox] (mu_z) at (10, 0.6) {$\mu_z$};
        \node[latentbox] (logsig_z) at (10, -0.6) {$\log\sigma_z^2$};
        
        \draw[arrow] (spread) -- (concat1);
        \draw[arrow] (embedding) -| (concat1);
        \draw[arrow] (concat1) -- (encmlp);
        \draw[arrow] (encmlp) -- (mu_z.west);
        \draw[arrow] (encmlp) -- (logsig_z.west);
        
        % 3. Latent Space
        \node[opcircle] (reparam) at (11.5, 0) {$\mathcal{N}$};
        \node[latentbox, fill=red!20, minimum height=1.2cm] (z) at (13.5, 0) {\textbf{Latent $z_t$} \\ $\in \mathbb{R}^3$};
        \node[above=0.3cm of reparam, font=\footnotesize\color{red!80}] {KL Pen.};
        
        \draw[arrow] (mu_z.east) -- (reparam);
        \draw[arrow] (logsig_z.east) -- (reparam);
        \draw[arrow] (reparam) -- (z);
        
        % 4. Decoder & 3-Heads
        \node[opcircle] (concat2) at (15.5, 0) {$\oplus$};
        \node[mlpbox] (decmlp) at (17.5, 0) {Shared Decoder \\ \footnotesize(Frozen in Stage B)};
        
        \draw[arrow, draw=orange!80, thick] (level.north) -- ++(0, 1.5) -| node[above, pos=0.2] {Level Re-injection (LevelScript)} (concat2.north);
        \draw[arrow, draw=orange!80, thick] (embedding.south) -- ++(0, -1.5) -| node[below, pos=0.25] {Currency Conditioning} (concat2.south);
        \draw[arrow] (z) -- (concat2);
        \draw[arrow] (concat2) -- (decmlp);
        
        \node[headbox] (loc) at (20, 0.8) {Loc $\mu_\theta$};
        \node[headbox] (scale) at (20, 0) {Scale $\lambda_\theta$};
        \node[headbox] (dof) at (20, -0.8) {DoF $\nu_\theta$};
        
        \draw[arrow] (decmlp.east) -- (loc.west);
        \draw[arrow] (decmlp.east) -- (scale.west);
        \draw[arrow] (decmlp.east) -- (dof.west);
        
        \node[opcircle, minimum size=0.8cm] (student) at (22, 0) {$T$};
        \node[inputbox, fill=green!10, draw=green!70] (output) at (24.5, 0) {Bond $P(t,T)$};
        \draw[arrow] (loc.east) -- (student);
        \draw[arrow] (scale.east) -- (student);
        \draw[arrow] (dof.east) -- (student);
        \draw[arrow] (student) -- (output);
        
        % 背景框 - Stage A
        \begin{scope}[on background layer]
            \node[draw=black!30, dashed, rounded corners, fill=gray!2, fit=(level)(embedding)(output)(mu_z), inner sep=0.4cm] (stageA_bg) {};
            \node[above=0.1cm of stageA_bg.north, font=\bfseries\Large] {Stage A: Manifold Learning (CVAEsT + LevelScript)};
        \end{scope}

        % ==========================================
        % STAGE B: Physics-Informed Neural SDE (下半场)
        % ==========================================
        
        % 5. Neural SDE (ParamNet)
        \node[mlpbox, fill=purple!5, draw=purple!70] (paramnet) at (13.5, -4) {ParamNet (Neural SDE) \\ \footnotesize(256$\times$3)};
        \draw[arrow, line width=1.5pt, draw=red!80] (z.south) -- node[right, font=\footnotesize] {Continuous Evolution} (paramnet.north);
        \draw[arrow, dashed, draw=orange!80] (embedding.south) |- node[below, pos=0.8, font=\footnotesize] {Conditioning} (paramnet.west);
        
        \node[physicsbox] (mup) at (10.5, -6) {Drift $\mu_{\mathbb{P}}$};
        \node[physicsbox] (sigma) at (13.5, -6) {Volatility $\Sigma$};
        \node[physicsbox] (lambda) at (16.5, -6) {Risk Premium $\lambda$};
        
        \draw[arrow] (paramnet.south) -- (mup.north);
        \draw[arrow] (paramnet.south) -- (sigma.north);
        \draw[arrow] (paramnet.south) -- (lambda.north);
        
        % 6. Measure Change
        \node[physicsbox, fill=yellow!20, draw=orange!100] (muq) at (13.5, -7.5) {Risk-Neutral Drift \\ $\mu_{\mathbb{Q}} = \mu_{\mathbb{P}} - \Sigma \lambda$};
        \draw[arrow] (mup.south) |- (muq.west);
        \draw[arrow] (lambda.south) |- (muq.east);
        \draw[arrow] (sigma.south) -- (muq.north);
        
        % 7. The Physics Bridge (Jacobians)
        \node[inputbox, fill=blue!10, draw=blue!80] (derivatives) at (19.5, -4) {Compute Derivatives \\ $\nabla_z D^{(P)}, \nabla^2_{zz} D^{(P)}$};
        \draw[dasharrow, draw=blue!80, line width=1pt] (decmlp.south) -- node[right, align=left, font=\footnotesize] {Differentiable\\Manifold} (derivatives.north);
        
        % 8. PDE Constraint (Final Loss)
        \node[pdebox] (pde) at (16.5, -9.5) {\textbf{No-Arbitrage PDE Constraint ($\mathcal{L}_{arb}$)} \\ \footnotesize $\mathcal{R}_{arb} = -\partial_\tau D + \nabla_z D^\top \mu_{\mathbb{Q}} + \frac{1}{2}\text{Tr}(\Sigma\Sigma^\top \nabla^2_{zz} D) - r D = 0$};
        
        \draw[arrow, line width=1pt, draw=red!80] (muq.south) |- (pde.west);
        \draw[arrow, line width=1pt, draw=red!80] (sigma.south) ++(0.5,0) |- ++(0, -1.5) -| (pde.north);
        \draw[arrow, line width=1pt, draw=blue!80] (derivatives.south) |- (pde.east);
        
        % 背景框 - Stage B
        \begin{scope}[on background layer]
            \node[draw=purple!40, dashed, rounded corners, fill=purple!2, fit=(paramnet)(mup)(lambda)(pde)(derivatives), inner sep=0.5cm] (stageB_bg) {};
            \node[below=0.1cm of stageB_bg.south, font=\bfseries\Large] {Stage B: Continuous-Time Dynamics \& Physical Gravity};
        \end{scope}
    \end{tikzpicture}
    }

    \caption{The Complete Physics-Informed Generative Architecture. \textbf{Stage A (Top):} The CVAEsT+LS isolates macroscopic curve shapes from absolute base rates (LevelScript) and utilizes a Student-t likelihood to absorb heavy-tailed anomalies, constructing a robust, low-dimensional continuous manifold $z_t$. \textbf{Stage B (Bottom):} A Neural SDE governs the temporal evolution of $z_t$. To satisfy theoretical pricing laws, the physical drift $\mu_{\mathbb{P}}$ is translated into the risk-neutral drift $\mu_{\mathbb{Q}}$ via the learned market price of risk $\lambda$. The entire system is end-to-end penalized by the No-Arbitrage PDE residual evaluated through the analytical derivatives of the frozen Stage A decoder.}
    \label{fig:full_architecture}
\end{figure*}

\begin{figure}[H]
  \centering
  \includegraphics[width=\textwidth]{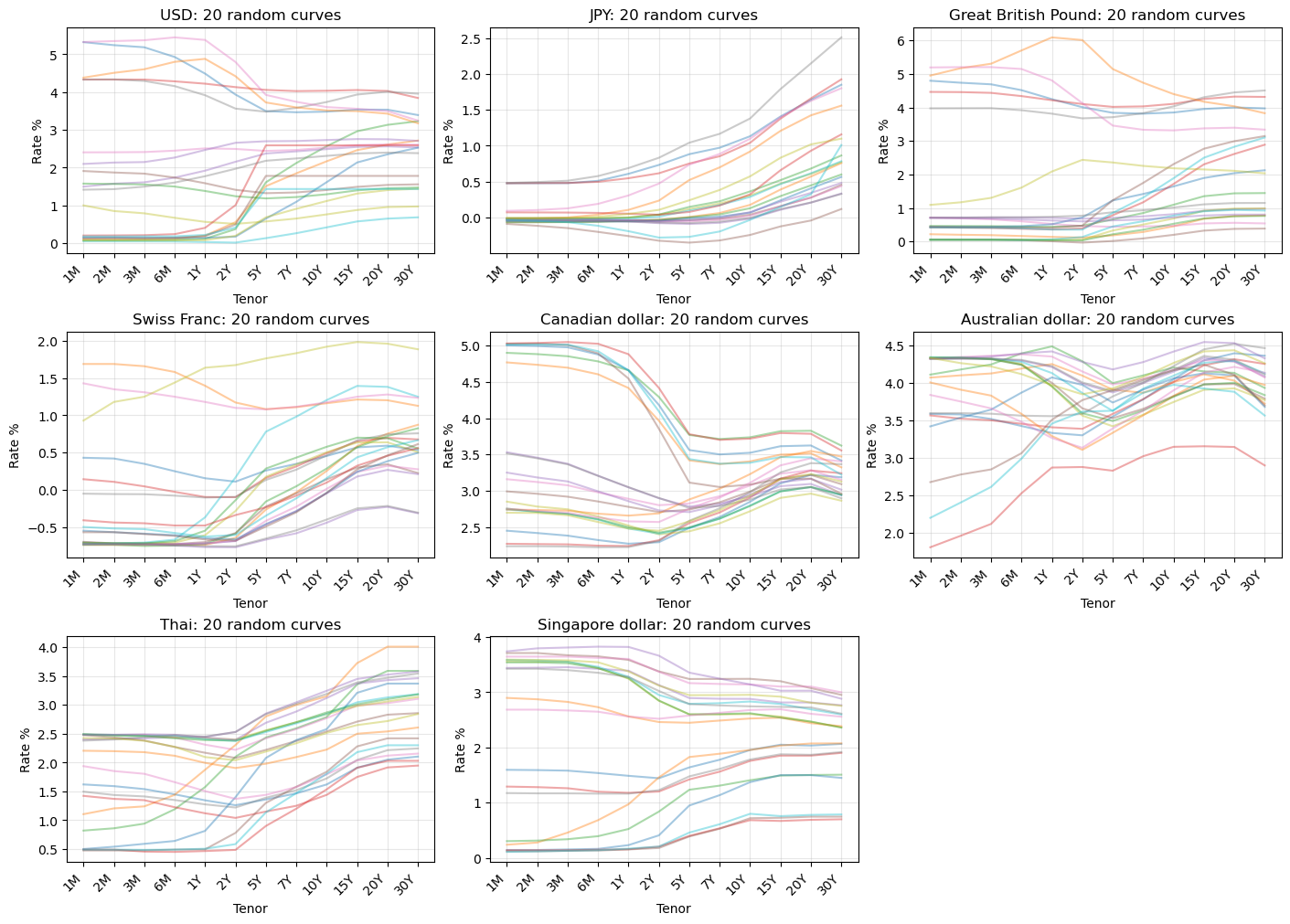}
  \caption{\textbf{Random samples of daily swap-rate curves across currencies}. For each currency (USD, JPY, GBP, CHF, CAD, AUD, THB, SGD), we plot 20 randomly selected daily curves on the standardized 12-tenor grid. Overlaid curves highlight within-currency variability in level, slope, and curvature, serving as a visual sanity check for cross-market comparability and data quality before constructing the VAE training panel.}
  \label{fig:daily_multicurrency}
\end{figure}

\begin{figure}[H]
  \centering
  \includegraphics[width=\textwidth]{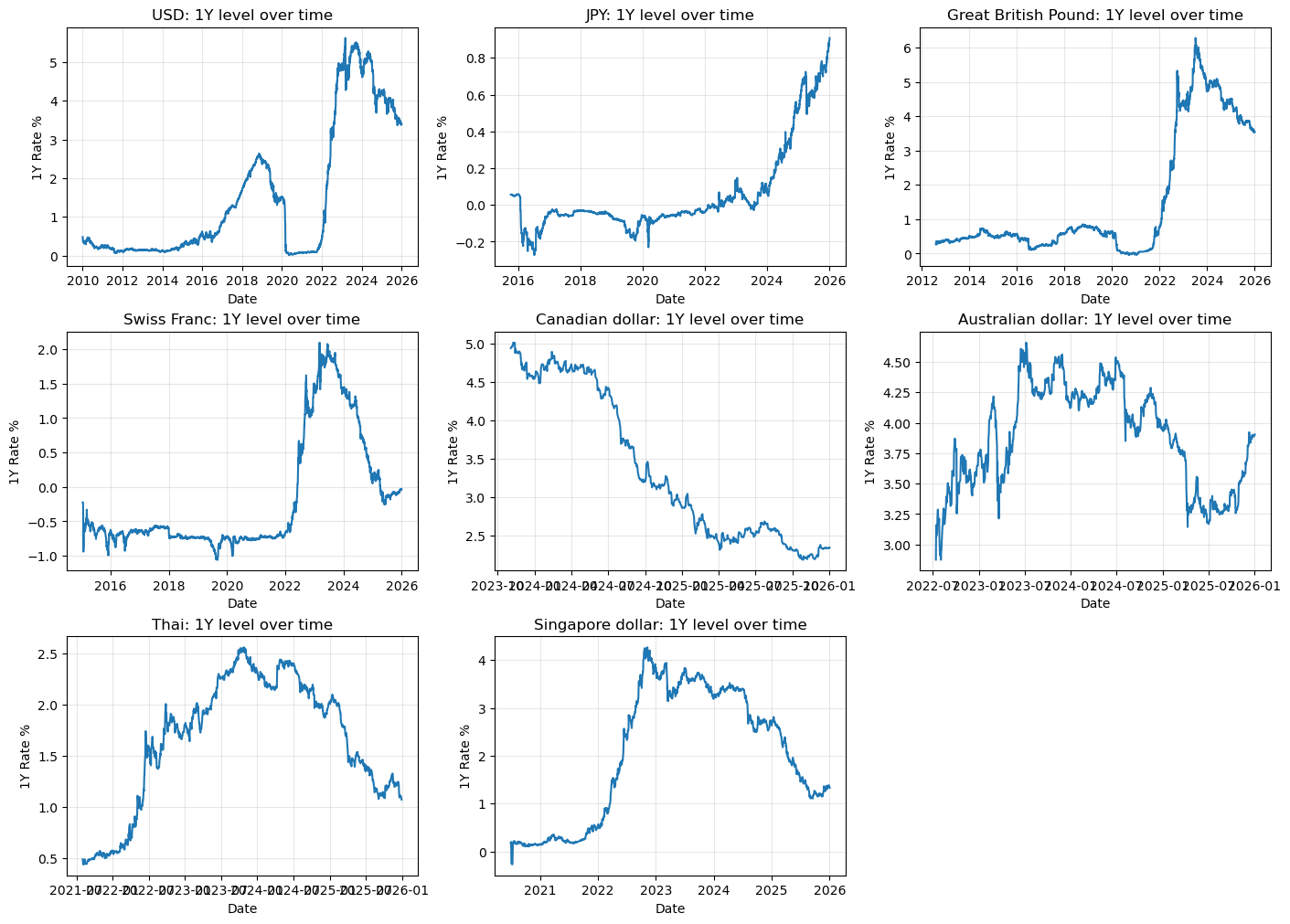}
  \caption{\textbf{1-year level dynamics across currencies}. The figure shows the $1$Y swap-rate time series for each currency in the final panel (USD, JPY, GBP, CHF, CAD, AUD, THB, SGD), plotted over the currency-specific post-start sample window. The $1$Y rate serves as the level anchor in our shape–level decomposition and illustrates cross-market regime variation that the VAE is designed to capture jointly with term-structure shape.}
  \label{fig:1y_multicurrency}
\end{figure}

\begin{figure}[H]
  \centering
  \includegraphics[width=0.6\textwidth]{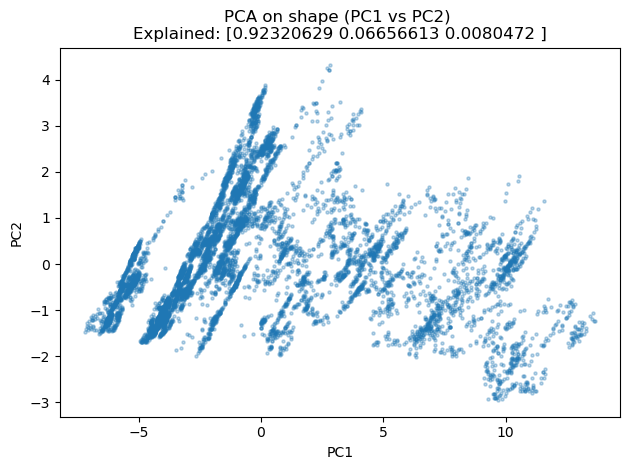}
  \caption{\textbf{PCA embedding of multi-currency curve shapes (PC$1$ vs PC$2$)}. Each point represents one daily curve observation after level removal (anchored at $1$Y) and robust scaling. The scatter plot shows the projection onto the first two principal components; the explained-variance ratios $[0.923, 0.067, 0.008]$ suggest that curve shape variability is largely two-dimensional, motivating a VAE with a $2$–$3$ dimensional latent space.}
  \label{fig:pca_multicurrency}
\end{figure}

As illustrated in the architecture diagram \ref{fig:full_architecture}, our framework consists of two deeply integrated components: Manifold Learning and Shape Decoupling (Stage A), and Physics-Constrained Continuous-Time Dynamics (Stage B). The primary objective of Stage A is to construct a robust, low-dimensional continuous manifold that captures pure macroeconomic shape dynamics. Given the differentiable latent manifold established in Stage A, Stage B governs the temporal evolution of $z_t$ by enforcing strict financial laws. We deploy a continuous-time Neural SDE (ParamNet) to learn the physical drift ($\mu_{\mathbb{P}}$) and volatility ($\Sigma$), while simultaneously extracting the dynamic market price of risk ($\lambda$). Subsequent sections explain the full picture of our pipeline.

\subsection{Stage A: Dimensionality Reduction via Swap-to-Bond Autoencoder}
Let $x_t \in \mathbb{R}^m$ denote the observed market data vector at time $t$. In this work, $x_t$ represents the vector of market swap rates for a fixed grid of tenors $\{\tau_1, \dots, \tau_m\}$. We learn an encoder-decoder pair where the decoder outputs the Zero-Coupon Bond (ZCB) price curve rather than reconstructing the inputs directly. The architecture is defined as:
\begin{equation}
    z_t = E_\psi(x_t), \quad \hat{P}(t,T) = D_\psi^{(P)}(z_t, \tau)
\end{equation}
where $z_t \in \mathbb{R}^d$ is the low-dimensional latent state ($d \ll m$), $\tau = T - t$ is the time-to-maturity, and $D_\psi^{(P)}: \mathbb{R}^d \times \mathbb{R}_+ \rightarrow (0, 1]$ is the bond-price decoder. To train Stage A, the decoded bond prices $\hat{P}(t,T)$ are mapped back to par swap rates $\hat{x}_t$ using the standard swap pricing formula:
\begin{equation}
    \hat{S}(t, T_n) = \frac{1 - \hat{P}(t,T_n)}{\sum_{i=1}^n \delta_i \hat{P}(t,T_i)}
\end{equation}
where $\delta_i$ represents the accrual period. The parameters are optimized to minimize the reconstruction error $||x_t - \hat{x}_t||^2$. After Stage A, the decoder $D_\psi^{(P)}$ is frozen (denoted as $D^{(P)}$), serving as a differentiable manifold map from latent states to the arbitrage-free discount curve.

\subsection{Stage B: Latent Diffusion and Measure Change}
To rigorously determine the intrinsic dimensionality of the yield curve variations after stripping the absolute levels, we perform a PCA on the preprocessed, robustly scaled multi-currency swap spreads anchored at the 1Y tenor. As illustrated in Figure \ref{fig:pca_multicurrency}, the eigenvalues drop precipitously: the first three principal components account for 92.32\%, 6.56\%, and 0.80\% of the total shape variance, respectively. Cumulatively, these three components capture over 99.68\% of the term structure's geometric deformations across all eight sovereign markets. This finding demonstrates the fact that only 2-3 factors are sufficient enough to determine the shape of a yield curve [\cite{bergeron2021variationalautoencodershandsoffapproach},  \cite{christensen2011affine}, \cite{lyashenko2024autoencoder}].

We follow similar methods used in the paper of Andreasen \cite{Andreasen2023Decoding}, where the no-arbitrage condition is first addressed with Neural SDEs. We modify this method by considering both $\mathbb{P}$ and $\mathbb{Q}$ measures and introduce a market risk factor $\lambda$ to facilitate the transition between the two measures $\mu_{\mathbb{Q}}(z)$ and $\mu_{\mathbb{P}}(z)$. As established in our theoretical review (Section \ref{sec:background}), the mathematical foundation for our method is rooted in Filipovi\'c's consistency theorem \cite{filipovic2001consistency}. 

This empirical decomposition provides a definitive statistical justification for setting the latent space dimension of our conditional generative framework to exactly three ($z \in \mathbb{R}^3$). Furthermore, the complex, filametary, and multi-branched geometric trajectories visible in the PC1-PC2 scatter plot suggest underlying non-linear regime shifts and sovereign-specific dynamics. While linear PCA successfully identifies the coordinate upper-bound, it lacks the expressive capacity to model these non-linear cross-currency topologies continuously, thereby fully motivating the deployment of our deep non-linear $CVAE_{sT}$ manifold learning architecture.

Given the latent time series $\{z_t\}$ extracted in Stage A, we model its evolution under the physical measure $\mathbb{P}$ to capture empirical dynamics, and transform it to the risk-neutral measure to enforce pricing consistency. We assume $z_t$ follows a stochastic differential equation (SDE):
\begin{equation}
    dz_t = \mu_\theta^{\mathbb{P}}(z_t)dt + \sigma_\theta(z_t)dW_t^{\mathbb{P}}
\end{equation}
where $\mu_\theta^{\mathbb{P}}$ and $\sigma_\theta$ are parameterized drift and diffusion functions (e.g., neural networks). From Girsanov's theorem, we introduce a market price of risk (MPR) process $\lambda_\phi(z_t)$ to define the dynamics used for pricing under the risk-neutral measure $\mathbb{Q}$:
\begin{equation}
    dW_t^{\mathbb{Q}} = dW_t^{\mathbb{P}} + \lambda_\phi(z_t)dt
\end{equation}
Substituting this into the physical SDE equation yields the risk-neutral dynamics:
\begin{equation}
    dz_t = \mu_{\theta, \phi}^{\mathbb{Q}}(z_t)dt + \sigma_\theta(z_t)dW_t^{\mathbb{Q}}
\end{equation}
where the risk-neutral drift is constrained by:
\begin{equation}
    \mu_{\theta, \phi}^{\mathbb{Q}}(z_t) = \mu_\theta^{\mathbb{P}}(z_t) - \sigma_\theta(z_t)\lambda_\phi(z_t)
\end{equation}

\subsection{Deriving the No-Arbitrage (PDE) Loss}
A fundamental condition of no-arbitrage is that the discounted bond price must be a martingale under $\mathbb{Q}$. Let $r_t = r_\theta(z_t)$ be the instantaneous short rate, modeled as a function of the latent state (or derived from the decoder gradient at $\tau \rightarrow 0$). The fundamental pricing equation implies:
\begin{equation}
    \mathbb{E}_t^{\mathbb{Q}}[dP(t,T)] = r_t P(t,T)dt
\end{equation}
Since the bond price is given directly by the decoder $P(t,T) = D^{(P)}(z_t, \tau)$, and noting that $d\tau = -dt$, we apply multi-dimensional It\^o's Lemma to $D^{(P)}$:
\begin{equation}
    dD^{(P)} = \underbrace{\left( -\partial_\tau D^{(P)} + \nabla_z D^{(P)\top}\mu_{\mathbb{Q}} + \frac{1}{2}Tr\left(\sigma_\theta\sigma_\theta^\top \nabla_{zz}^2 D^{(P)}\right) \right)dt}_{\text{Drift under } \mathbb{Q}} + \nabla_z D^{(P)}\sigma_\theta dW_t^{\mathbb{Q}}
\end{equation}
Note that the term $Tr(\cdot)$ comes from the application of convexity adjustment. For the discounted bond price to be a martingale, the drift of the bond price must equal the short rate times the price:
\begin{equation}
    \text{Drift under } \mathbb{Q} = r_\theta(z_t)D^{(P)}(z_t, \tau)
\end{equation}
Equating the terms yields the partial differential equation (PDE) that our decoder and latent dynamics must satisfy:
\begin{equation}
    \mathcal{R}_{arb}(z,\tau) := -\partial_\tau D^{(P)} + \nabla_z D^{(P)\top}\mu_{\theta,\phi}^{\mathbb{Q}} + \frac{1}{2}Tr\left(\sigma_\theta\sigma_\theta^\top \nabla_{zz}^2 D^{(P)}\right) - r_\theta D^{(P)} = 0
\label{eq:pde}
\end{equation}
This equation is significantly simpler than the differential equations required when decoding forward rates, as it avoids the integral term.

\subsection{Training Objective}
Stage B is trained to minimize a composite loss function that balances physical data likelihood with the no-arbitrage constraint. To ensure stable gradient descent and remove the measurements effect from different currencies, we use a Sharpe-ratio-style normalization. We scale the raw residual $\mathcal{R}_{arb}$ by the instantaneous volatility of the bond price, $||\nabla_z D^{(P)\top}\sigma_\theta||_{\text{Euclidean}}$:
\begin{equation}
    \mathcal{L}_{arb}^{LN} = \mathbb{E}_{z,\tau} \left[ \left( \frac{\mathcal{R}_{arb}(z,\tau)}{\sqrt{||\nabla_z D^{(P)\top}\sigma_\theta||_{\text{Euclidean}}^2 + \epsilon}} \right)^2 \right]
\end{equation}
In this way, our target loss function is scale-free, and this prevents the neural network from "cheating" by optimizing only the short end of the curve while ignoring the long end. In addition, the $\epsilon$ term is introduced to avoid division by zero. We optimize the parameters $\{\theta, \phi\}$ (dynamics and risk price) to minimize:
\begin{equation}
    \mathcal{L} = \mathcal{L}_{data}(\mu_{\mathbb{P}}, \sigma) + \beta\mathcal{L}_{arb}^{LN}(\mu_{\mathbb{Q}}, \sigma, r) + \gamma||\lambda_\phi||^2
\end{equation}
where $\mathcal{L}_{data}$ is the negative log-likelihood of the observed latent transitions under $\mathbb{P}$, and the final term regularizes the market price of risk. We build a Multi-Layer Perceptron (MLP) mapping $ParamNet(z) \mapsto (\mu_{\mathbb{P}}(z), \Sigma(z), \lambda(z))$, yielding $\mu_{\mathbb{Q}}(z) = \mu_{\mathbb{P}}(z) - \Sigma(z)\lambda(z)$.

\subsection{Discrete-Time Realization of the Neural SDE}
\label{sec:discretization}
While the theoretical framework models the latent yield curve dynamics as a continuous-time Stochastic Differential Equation, empirical data is observed at discrete daily intervals, and neural network tensor operations require a discretized integration scheme. To bridge the continuous theoretical physics with our PyTorch implementation, we employ the Euler-Maruyama discretization scheme. Let $\Delta t$ denote the discrete time step. Given that our dataset consists of daily OIS swap rates, we calibrate the time step to the annual trading calendar, setting $\Delta t = 1/252$. The continuous physical transition is thereby discretized as follows:
\begin{equation}
    z_{t+\Delta t} = z_t + \mu_{\mathbb{P}}(z_t)\Delta t + \Sigma(z_t)\sqrt{\Delta t}\epsilon_{t+1}, \quad \epsilon_{t+1} \sim \mathcal{N}(0, I_d)
\end{equation}
where $\epsilon_{t+1}$ is a standard Gaussian noise vector sampled at each forward step. Consequently, the physical data likelihood $\mathcal{L}_{data}$ utilized in the training objective is explicitly computed as the negative log-likelihood of the Gaussian transition probability:
\begin{equation}
    z_{t+\Delta t} | z_t \sim \mathcal{N}\left(z_t + \mu_{\mathbb{P}}(z_t)\Delta t, \Sigma(z_t)\Sigma(z_t)^\top \Delta t\right)
\end{equation}
Similarly, to enforce the No-Arbitrage PDE constraint across simulated paths, the risk-neutral dynamics are discretized using the learned market price of risk $\lambda(z_t)$:
\begin{equation}
    z_{t+\Delta t}^{\mathbb{Q}} = z_t + \mu_{\mathbb{Q}}(z_t)\Delta t + \Sigma(z_t)\sqrt{\Delta t}\epsilon_{t+1}^{\mathbb{Q}}
\end{equation}
By strictly anchoring the forward simulations to $\Delta t = 1/252$ and utilizing the Euler-Maruyama method, we ensure that the theoretical drift and volatility are scaled correctly in the annual domain, eliminating temporal discretization misalignment.

\subsection{Exact Computation of the PDE Penalty via Automatic Differentiation}
A central computational challenge in our framework is the evaluation of the No-Arbitrage PDE residual (Equation \ref{eq:pde}), specifically the convexity adjustment term which requires computing the trace of the Hessian matrix: $\frac{1}{2}Tr(\Sigma\Sigma^\top \nabla_{zz}^2 D^{(P)})$. 

In traditional quantitative finance, such high-dimensional second-order derivatives are typically approximated using Finite Difference Methods (FDM). However, applying FDM to deep neural networks introduces severe truncation errors, numerical instability, and an intractable computational overhead (the curse of dimensionality). To circumvent these classical bottlenecks and ensure complete mathematical transparency, our framework eschews numerical approximation entirely in favor of Automatic Differentiation (Autograd). Unlike numerical approximations that suffer from truncation errors, we employ reverse-mode automatic differentiation to compute exact gradients to machine precision \cite{baydin2018automatic}.

Let $D_k^{(P)}$ denote the decoded bond price for the $k$-th tenor. The first-order Jacobian vector is computed via a single backward pass:
\begin{equation}
    J_k = \nabla_z D_k^{(P)} = \left[ \frac{\partial D_k^{(P)}}{\partial z_1}, \frac{\partial D_k^{(P)}}{\partial z_2}, \dots, \frac{\partial D_k^{(P)}}{\partial z_d} \right]^\top
\end{equation}
To obtain the Hessian required for the PDE penalty without explicitly materializing the full $d \times d \times d$ tensor for all batch elements (which would severely bottleneck memory), we employ the Hessian-vector product trick permitted by the $C^\infty$ smoothness of our chosen Softplus activations. The exact $(i, j)$-th element of the Hessian matrix is derived as:
\begin{equation}
    [\nabla_{zz}^2 D_k^{(P)}]_{i,j} = \frac{\partial^2 D_k^{(P)}}{\partial z_i \partial z_j}
\end{equation}

Specifically, explicitly materializing the full Hessian tensor scales as $\mathcal{O}(d^2)$ in memory, which is intractable for large batch sizes. By leveraging the Pearlmutter trick \cite{pearlmutter1994fast}, we evaluate the Hessian-vector products (HVP) directly via $\nabla_z(\langle \nabla_z D^{(P)}, v \rangle)$ using consecutive reverse-mode graph traversals. Combined with stochastic trace estimators introduced by Hutchinson \cite{hutchinson1989stochastic}, this permits the evaluation of the PDE convexity adjustment strictly in $\mathcal{O}(d)$ memory complexity. This architectural choice is precisely what enables the Neural SDE to maintain stable gradient descent and successfully learn the reflective boundary dynamics observed in constrained regimes like the JPY.

% ============ Section5: Advanced Theoretical Benchmarks ================

\section{Advanced Theoretical Benchmarks}
\label{sec:benchmarks}

While our empirical analysis evaluates the proposed architecture against classical baselines (e.g., Standard HJM and Unconstrained VAR), it is imperative to acknowledge the emergence of other highly sophisticated generative frameworks in recent literature. Although a full empirical replication of these specialized architectures falls outside the scope of the current study, we formalize two advanced theoretical benchmarks below. These models represent elegant alternative approaches to combining deep learning with fixed-income pricing and serve as natural and robust comparators for future research.

\subsection{Benchmark 1: Dynamic Nelson-Siegel (DNS) Refined in Autoencoder Latent Space ($\mathbb{P}$-measure)}
This benchmark, originally introduced in Diebold and Li \cite{diebold2006forecasting}, isolates the effect of a learned curve manifold by keeping the temporal dynamics intentionally simple (linear autoregression), while the representation of admissible curve shapes is data-driven. Let $y_t \in \mathbb{R}^m$ denote the observed term-structure vector at time $t$. In this framework, an autoencoder is trained to learn a low-dimensional representation: $z_t = E(y_t)$ and $\hat{y}_t = D(z_t)$.

After training, a DNS-style forecasting model is defined by replacing the classical Nelson-Siegel factors with the latent state $z_t$ and imposing simple autoregressive dynamics under the real-world measure $\mathbb{P}$:
\begin{equation}
    z_{t+h} = \alpha + \Phi z_t + \epsilon_{t+h}, \quad \epsilon_{t+h} \sim \mathcal{N}(0, \Sigma)
\end{equation}
where $h$ is the forecast horizon, $\Phi \in \mathbb{R}^{d \times d}$ is the transition matrix (diagonal for an AR(1)-per-factor or full for a VAR(1) process), and $\Sigma$ is the innovation covariance. The predictive distribution of future curves is obtained by sampling $z_{t+h} | z_t$ from the latent model and decoding: $\hat{y}_{t+h} = D(z_{t+h})$.

\subsection{Benchmark 2: Forward-Rate AEMM with Decoder-Jacobian Volatility Basis ($\mathbb{Q}$-measure)}
An alternative to our Neural SDE penalty approach is brought up by Sokol \cite{sokol2022autoencoder}, to construct an Arbitrage-Free Energy-Based/Manifold Model (AEMM) that explicitly uses the geometry of the neural network. Let $f_t(\tau)$ denote the instantaneous forward curve at time $t$. An autoencoder is trained on historical forward curves, obtaining a decoder $\hat{f}(\tau; z) = D(z)(\tau)$ and an encoder $z(f) = E(f)$.

The forward-rate AEMM specifies an HJM-style risk-neutral ($\mathbb{Q}$) dynamics where the volatility basis is explicitly given by the local geometry of the decoder at the current state. For each latent coordinate $k=1, \dots, d$, the state-dependent basis function (a tangent direction) is defined as:
\begin{equation}
    b_k(\tau; f_t) = \left. \frac{\partial f(\tau; z)}{\partial z_k} \right|_{z=E(f_t)}
\end{equation}
The forward-curve evolution under $\mathbb{Q}$ is then modeled as:
\begin{equation}
    df_t(\tau) = \mu_t(\tau)dt + \sum_{k=1}^d \sigma_k(t)b_k(\tau; f_t)dW_{k,t}^{\mathbb{Q}}
\end{equation}
where $\sigma_k(t)$ controls the time-varying magnitude of factor $k$, and $\{W_k^{\mathbb{Q}}\}_{k=1}^d$ are independent Brownian motions. The drift $\mu_t(\tau)$ is deterministically set to satisfy the HJM no-arbitrage restriction. To prevent long-horizon drift or discretization errors from pushing trajectories off the learned manifold during simulation, a projection (re-encoding) step is strictly applied at each time increment:
\begin{equation}
    \tilde{f}_{t+\Delta t} \text{ (SDE step)} \rightarrow z_{t+\Delta t} = E(\tilde{f}_{t+\Delta t}), \quad f_{t+\Delta t} = D(z_{t+\Delta t})
\end{equation}
This benchmark gracefully preserves the classical risk-neutral framework while replacing hand-crafted factor loadings with a data-driven, state-dependent volatility basis derived from the autoencoder's Jacobian.

\section{Experimental Results}
\label{sec:results}

\subsection{Stage A: Yield Curve Learning and Ablation Study}
We first evaluate the representation quality of Stage A by reconstruction error on the validation set. Let $x_t \in \mathbb{R}^m$ be the scaled forward-rate vector on a dense maturity grid and $\hat{x}_t$ its reconstruction. We report the daily RMSE, which summarizes reconstruction accuracy across the full curve at each date. A comparative analysis of the RMSE across the VAE and CVAE models reveals critical insights into the benefits of conditional modeling.

\begin{figure}[htbp]
    \centering
    % 第一个子图
    \begin{subfigure}[b]{0.9\textwidth}
        \centering
        \includegraphics[width=\textwidth]{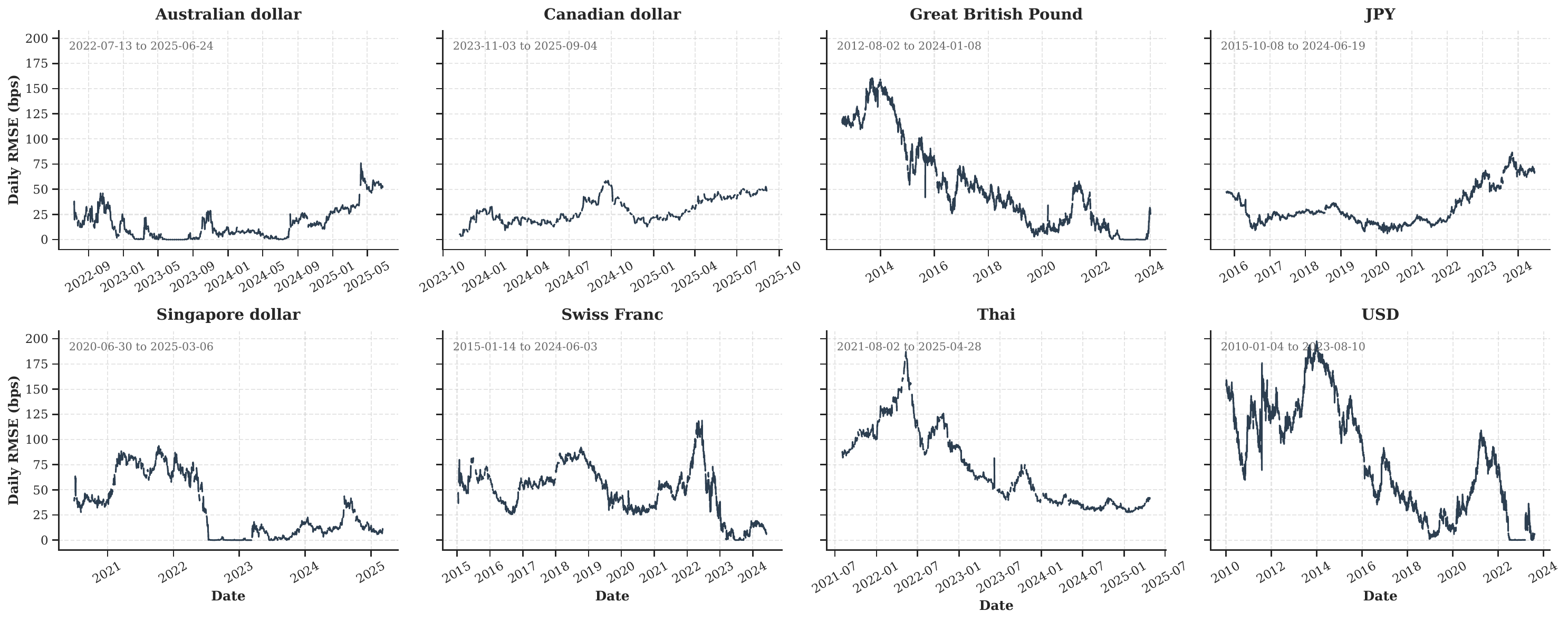}
        \caption{In-Sample daily RMSE Performance of the model VAEsT}
        \label{fig:new_vae_in}
    \end{subfigure}
    
    \vspace{1em} % 增加两个图之间的间距
    
    % 第二个子图
    \begin{subfigure}[b]{0.9\textwidth}
        \centering
        \includegraphics[width=\textwidth]{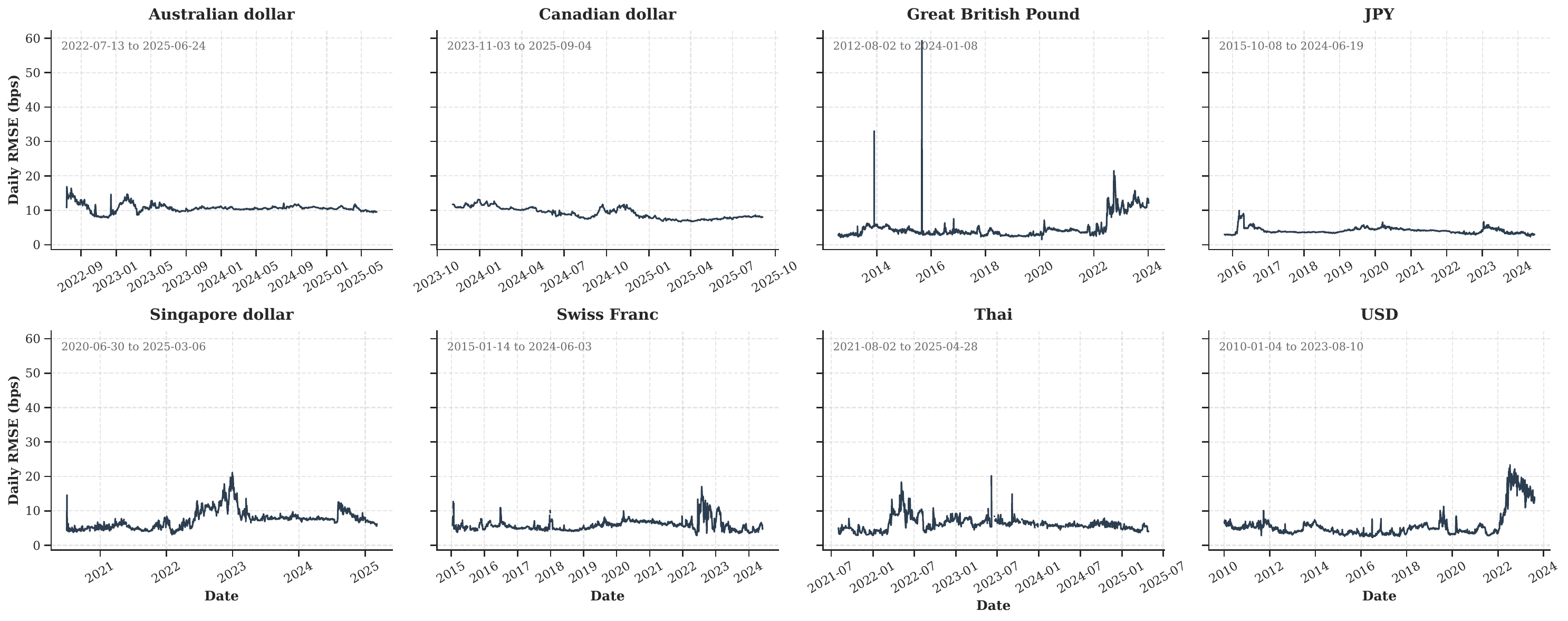}
        \caption{In-Sample daily RMSE Performance of the model CVAE$_{sT}$+LS}
        \label{fig:new_cvae_in}
    \end{subfigure}
    
    \caption{Performance evaluation of the models. (a) VAEs+StudentT; (b) CVAEs + StudentT + Level Script}
    \label{fig:new_performance_insample}
\end{figure}

\begin{figure}[htbp]
    \centering
    % 第一个子图
    \begin{subfigure}[b]{\textwidth}
        \centering
        \includegraphics[width=0.9\textwidth]{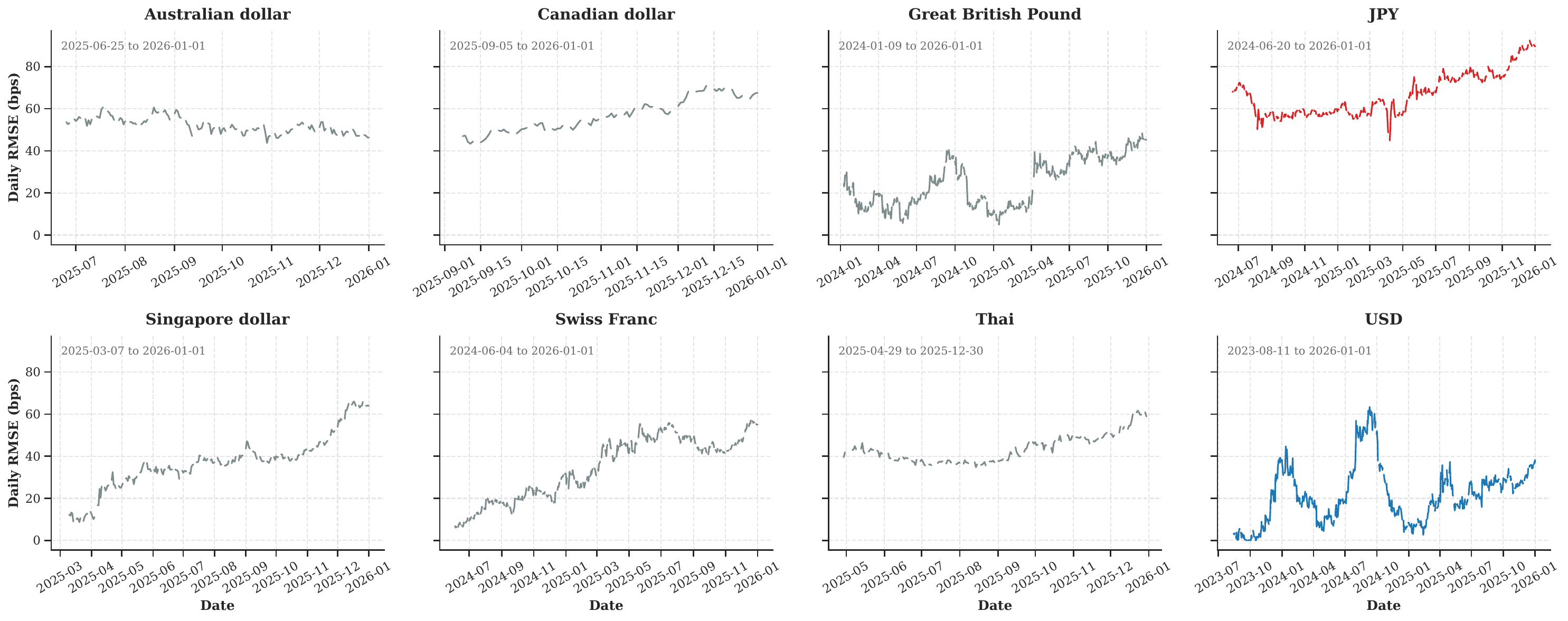}
        \caption{Out-of-Sample daily RMSE Performance of the model VAEsT}
        \label{fig:new_vae_oos}
    \end{subfigure}
    
    \vspace{1em} % 增加两个图之间的间距
    
    % 第二个子图
    \begin{subfigure}[b]{\textwidth}
        \centering
        \includegraphics[width=0.9\textwidth]{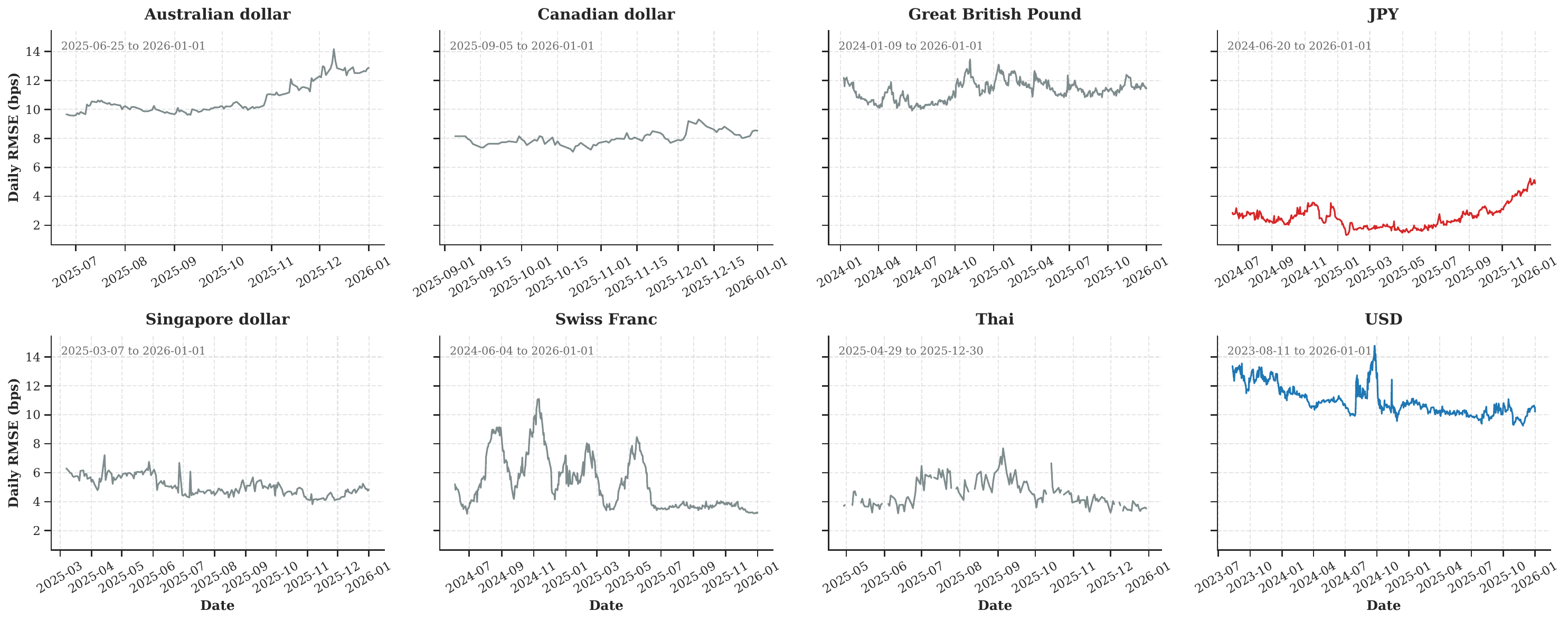}
        \caption{Out-of-Sample daily RMSE Performance of the model CVAE$_{sT}$+LS}
        \label{fig:new_cvae_oos}
    \end{subfigure}
    
    \caption{Performance evaluation of the models. (a) VAEs + StudentT; (b) CVAEs + StudentT + Level Script.}
    \label{fig:new_performance_oossample}
\end{figure}

\textbf{Performance Enhancement with the Proposed CVAE$_{sT}$+LS:}
Figure \ref{fig:new_performance_insample} illustrates the empirical performance of the proposed $\text{CVAE}_{sT}$+LS framework. In stark contrast to the standard VAE results presented in Figure \ref{fig:new_performance_oossample}, the $\text{CVAE}_{sT}$+LS architecture achieves an order-of-magnitude reduction in reconstruction error in both in-sample and out-of-sample data. 

The daily RMSE trajectories (Figure \ref{fig:new_performance_insample} and \ref{fig:new_performance_oossample}) are remarkably flat and stable on both in-sample and out-of-sample datasets of the model CVAE$_{sT}$+LS, with most currencies maintaining an error profile between 2 and 14 bps, effectively eliminating the volatile error spikes observed in the baseline model. Specifically, the model’s ability to anchor the curve at the 1Y tenor (as seen in the tenor-wise RMSE profile, right column) proves the effectiveness of the LevelScript mechanism in stripping absolute level bias before the generative process. 

Most importantly, the JPY reconstruction—which previously exhibited erratic behavior in the standard VAE—now demonstrates structural stability. This confirms that by incorporating both heavy-tailed likelihoods (Student-$t$) and a physical level anchor (LevelScript), the model successfully disentangles global macroeconomic shape dynamics from idiosyncratic sovereign levels. The result is a unified, continuous-time generative engine that preserves structural integrity across the entire term structure, regardless of the sovereign regime or interest rate environment.

\textbf{Ablation Study Insights:} Table \ref{tab:ablation} presents a comprehensive ablation study decomposing the out-of-sample RMSE. 
\begin{enumerate}
    \item \textit{The Baseline Failure:} Traditional PCA suffers from massive structural drift, while the standard Gaussian VAE and CVAE models, though vastly superior to PCA, still exhibit dangerously high errors during volatile macro regimes. This highlights that relying solely on Gaussian assumptions is fatal during aggressive monetary policy shifts.
    \item \textit{The Heavy-Tail Breakthrough:} When the Student-$t$ distribution is introduced, the Mean Currency RMSE plummets from 243.85 bps to just 12.02 bps. This confirms that the Student-$t$ distribution effectively absorbs idiosyncratic market shocks and fat-tailed risks, preventing the latent manifold from collapsing under extreme stress.
    \item \textit{The Physics Anchor:} The introduction of the LevelScript module provides a distinct type of structural improvement at the highly convex long-end of the curve. By dynamically stripping the absolute rate level, LevelScript acts as a physical anchor.
    \item \textit{The Ultimate Synergy:} The combination of Student-$t$ dynamics and the LevelScript anchor is highly synergistic. The proposed $\text{CVAE}_{sT} + \text{LS}$ model achieves an exceptionally low Mean Tenor RMSE of 6.58 basis points. It provides exceptional accuracy at the critical belly (1Y at 2.89 bps) and the highly sensitive long end (30Y at 5.13 bps), demonstrating its compliance with the stringent No-Arbitrage PDE convexity conditions seamlessly across both aggressive hiking environments (USD) and yield-curve-controlled environments (JPY).
\end{enumerate}

\subsection{Latent Dimensional Comparison and Disentanglement}
\label{sec:disentanglement}

\begin{figure*}[htbp]
    \centering
    \includegraphics[width=0.9\textwidth]{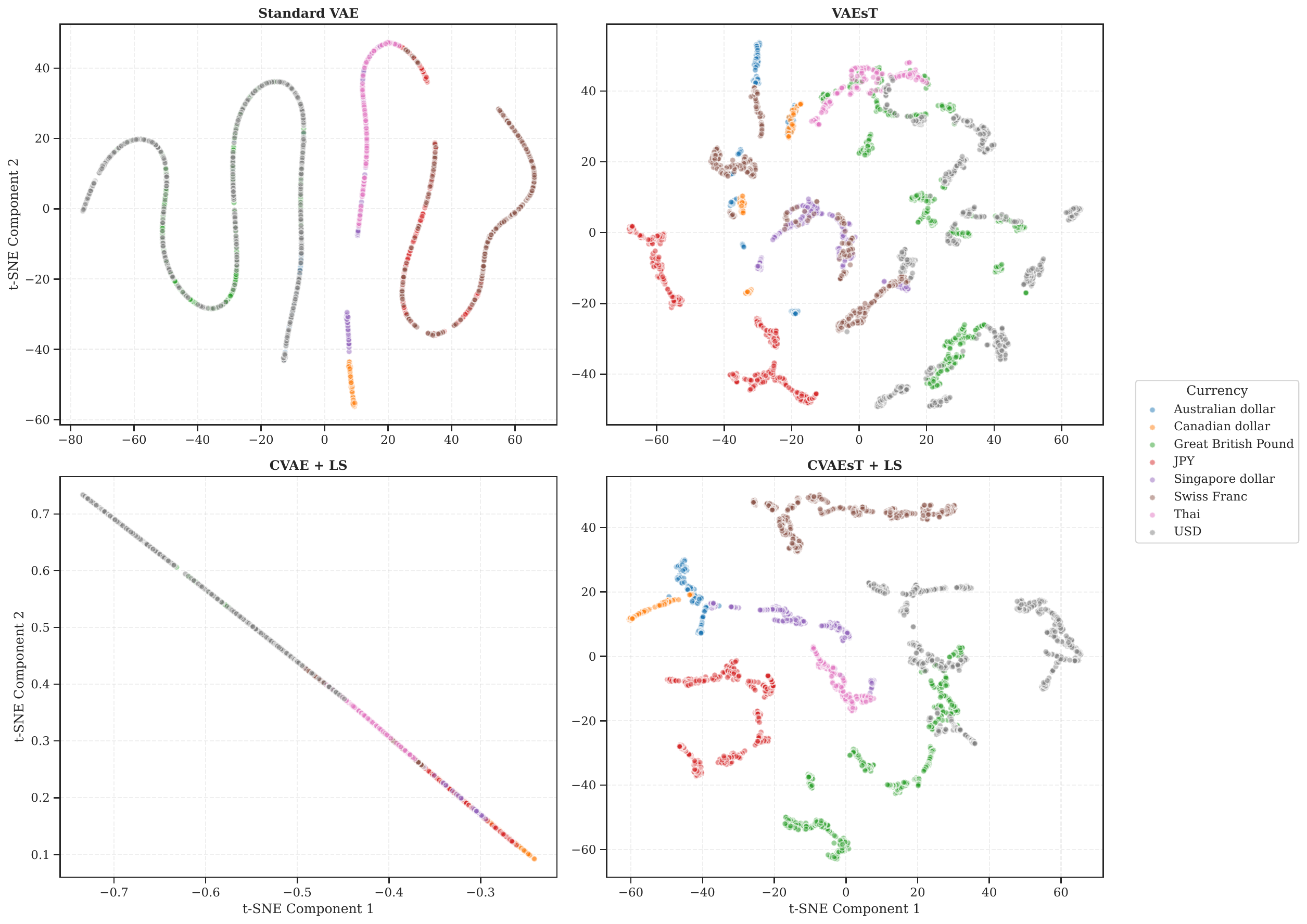}
    \caption{Evolution of Latent Manifold Disentanglement. From (a) to (d), we illustrate the transition from currency-specific clustering (baseline VAE) to a unified, disentangled latent manifold ($\text{CVAE}_{sT}$+LS). Each color represents a unique sovereign currency. The dissolution of distinct clusters in (d) confirms the successful decoupling of shared global shape factors from currency-specific base-rate biases.}
    \label{fig:latent_manifold}
\end{figure*}

Figure \ref{fig:latent_manifold} illustrates the evolution of latent feature disentanglement across four architectural stages. For all four models, we use Optuna to conduct hyperparameter-tuning, resulting in the dimensions where the lowest RMSE values lie \cite{akiba2019optunanextgenerationhyperparameteroptimization}. The baseline models (Fig. \ref{fig:latent_manifold}a, b) exhibit strong currency-specific clustering, indicating that the latent capacity is predominantly occupied by memorizing disparate absolute interest rate levels rather than learning shared macroeconomic shape dynamics. 

The transition to the CVAE+LS framework (Fig. \ref{fig:latent_manifold}c) demonstrates the potent effect of the LevelScript anchor. By stripping the absolute base rate, the model undergoes a dimensional collapse, forcing the latent representation onto a near-1D manifold. While this enforces strict structural invariance, it potentially sacrifices the representation of complex, heavy-tailed market configurations. 

Our proposed $\text{CVAE}_{sT}$+LS model (Fig. \ref{fig:latent_manifold}d) achieves the optimal balance. By combining the LevelScript anchor with a Student-$t$ likelihood, the model successfully avoids both the currency-biased clustering of the baseline VAEs and the over-constrained collapse of the CVAE+LS. The resulting latent space is uniformly distributed and cross-currency invariant, demonstrating that the model has internalized a universal ``shape language.'' The absence of distinct currency clusters—even for the JPY and USD—serves as empirical proof that our framework has successfully disentangled global macroeconomic risk factors from idiosyncratic base-rate biases.

\subsection{FinQ-VAEs Performance and Limitations}
The FinQ-VAE baseline provides a fascinating empirical look into the mechanics of hierarchical latent quantization. By forcing the network's subsequent layers ($L_0, L_1, L_2$) to explicitly fit the residual errors at designated anchor nodes (from 1Y to 30Y), the cascade architecture successfully forces the long end of the yield curve into mathematical alignment. This is clearly evidenced by the steady decline in error towards the 30Y tenor in figure \ref{fig:finq_daily}. The quantization acts as a series of structural "magnets," preventing the long-end predictions from drifting into mathematically unsound territory.

\begin{figure}[H]
  \centering
    \includegraphics[width=0.9\textwidth]{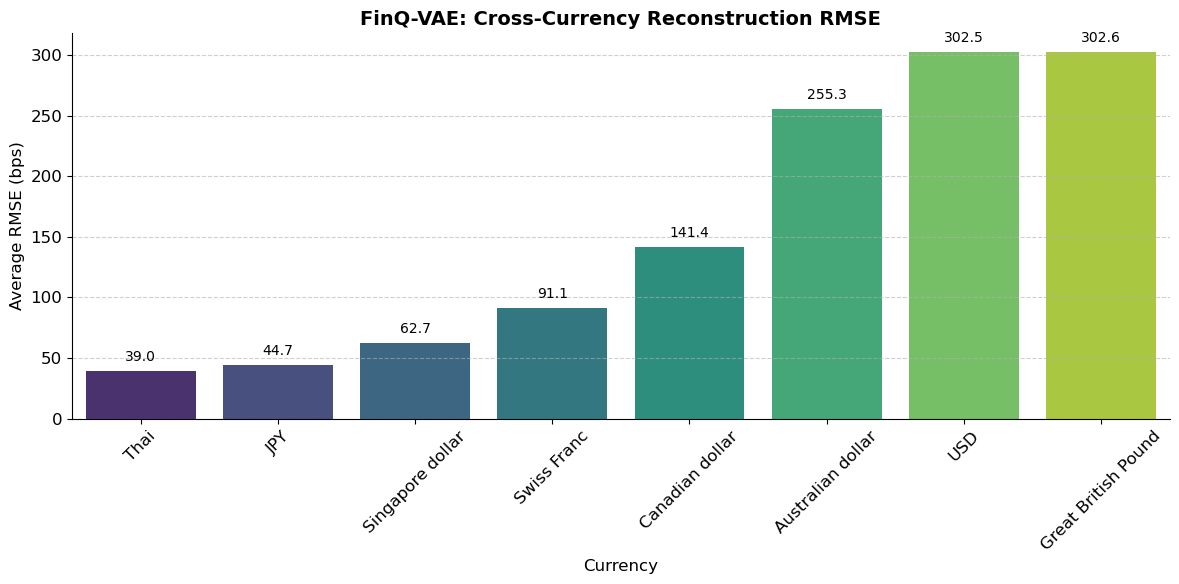}
  \caption{\textbf{Cross-Currency Average RMSE}: Average out-of-sample reconstruction error grouped by sovereign currency. The architecture demonstrates highly polarized performance. It achieves strong reconstruction accuracy for structurally stable or lower-yield regimes such as Thailand (39.0 bps) and Japan (44.7 bps), while suffering severe performance degradation in higher-magnitude or more volatile regimes, notably the USD (302.5 bps) and GBP (302.6 bps).}
  \label{fig:finq_currency}
\end{figure}

\begin{figure}[H]
  \centering
    \includegraphics[width=0.9\textwidth]{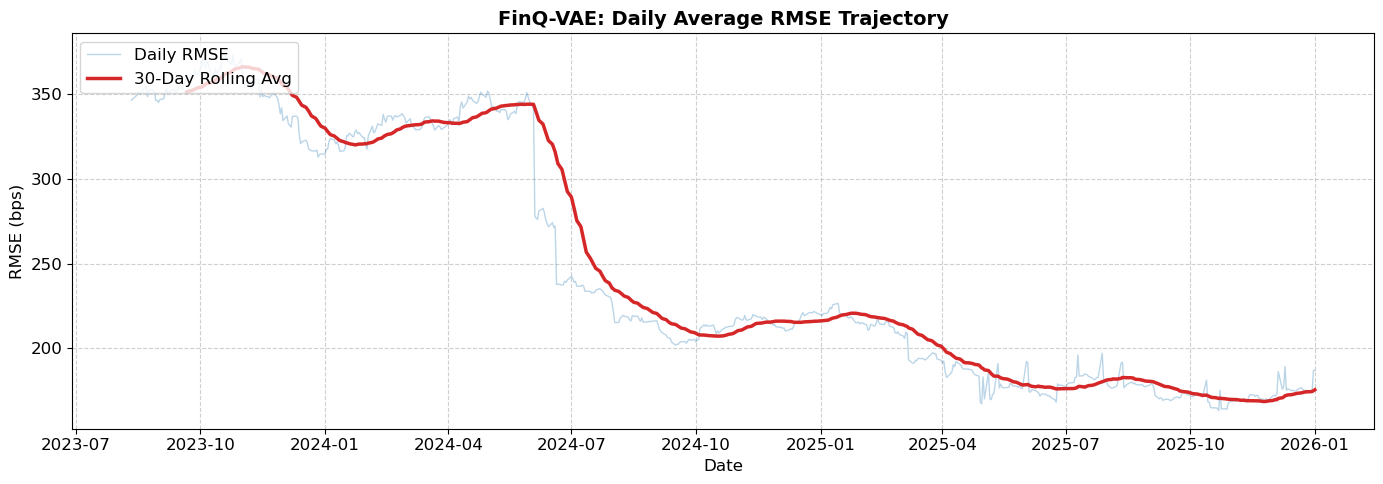}
  \caption{\textbf{Daily RMSE Trajectory (Rolling 30-Day Average)}: Chronological evolution of the out-of-sample RMSE across all currencies from mid-2023 through early 2026. The trajectory reveals a significant structural break and error reduction occurring around mid-2024. Following this regime shift, the model stabilizes its reconstruction performance, steadily trending downward toward the 170 bps range by the end of the out-of-sample period.}
  \label{fig:finq_daily}
\end{figure}

However, this evaluation exposes a critical structural vulnerability that perfectly validates the necessity of explicit level-stripping and conditionality. The model's errors at the short end (1M to 6M) remain disproportionately high, and its cross-currency performance is drastically inconsistent in Figure \ref{fig:finq_currency}. Because this specific FinQ-VAE implementation lacks a deterministic macroeconomic anchor (such as explicitly injecting the LIBOR or FFR), it attempts to use its quantized latent capacity to memorize absolute baseline heights rather than pure curve shapes. This explains the massive 300+ bps errors for the USD and GBP; the network simply cannot span the vast absolute yield gaps between a high-rate US environment and a low-rate Japanese environment using bounded, quantized latent spaces alone. The model learns to fit the localized "bumps" of the curve but fundamentally fails to anchor the curve in the correct absolute macroeconomic reality.

\subsection{Stage B: No-Arbitrage Training Signal and Dynamic SDE Evaluation}
The mathematical core of Stage B is the PDE residual, which strictly requires computing both the Jacobian $\nabla_z D^{(P)}$ and the trace of the Hessian $\nabla_{zz}^2 D^{(P)}$. A Gaussian decoder penalizes outliers quadratically. During market shocks, it violently contorts the latent space to fit the anomaly, creating "wrinkles" and sharp cliffs in the manifold. When Stage B calculates the second derivative (Hessian) at these cliffs, the convexity adjustment term $\frac{1}{2}Tr(\sigma_\theta\sigma_\theta^\top \nabla_{zz}^2 D^{(P)})$ produces massive numerical spikes, destabilizing the PDE loss. The Student-$t$ decoder absorbs market shocks in the observation space using heavy tails, leaving the latent manifold smooth and continuous. This guarantees well-behaved derivatives, making the PDE optimizer's job exponentially easier.

By offloading the baseline yield levels to the currency embedding $c$, the CVAE compresses $z_t$ into a ``universal yield curve language.'' The SDE can now learn a unified, continuous vector field for $\mu_{\mathbb{P}}(z_t)$ that applies globally, making the physical transition likelihood robust.

\begin{figure}[htbp]
    \centering

    \begin{subfigure}{0.9\textwidth}
        \centering
        \includegraphics[width=0.9\linewidth]{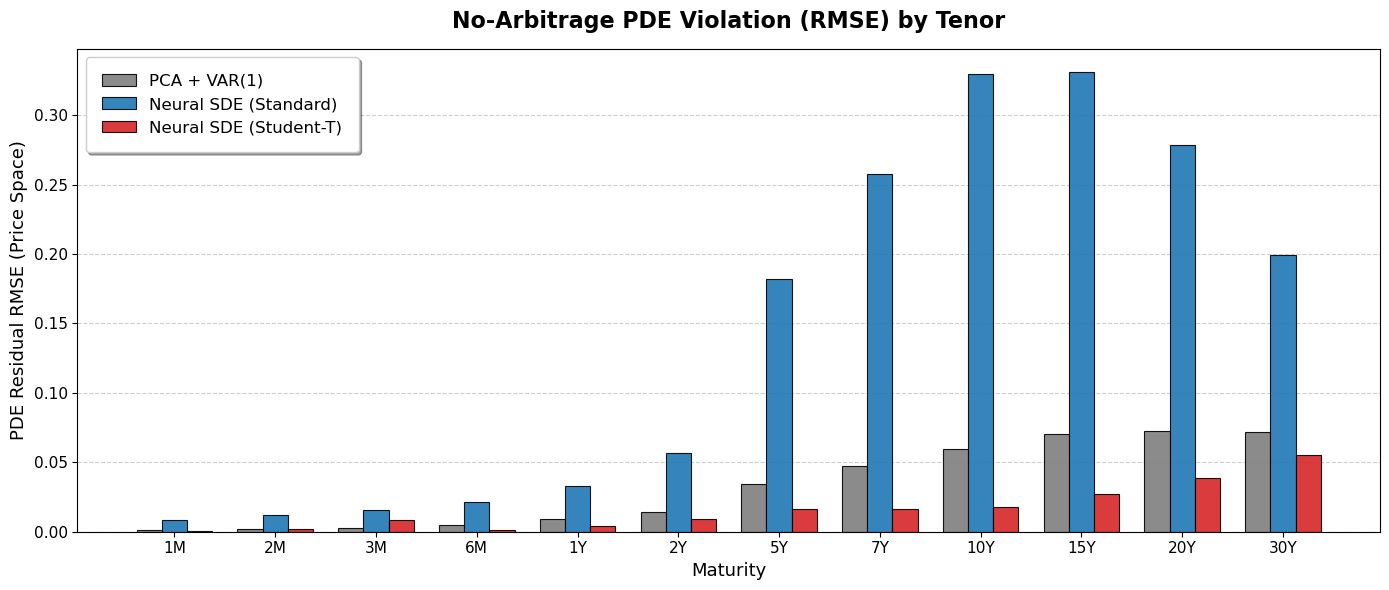}
        \caption{\textbf{No-Arbitrage PDE Violation by Tenor}: Evaluation of physical constraint satisfaction. The Standard SDE exhibits massive arbitrage violations at the mid-to-long tenors (5Y-30Y) due to its inability to accommodate extreme convexity under rigid Gaussian assumptions. In contrast, the heavy-tailed Student-t SDE strictly bounds the PDE residuals across all tenors, proving it successfully absorbs tail-risk shocks without breaking fundamental no-arbitrage pricing laws.}
    \end{subfigure}
    \caption{No-arbitrage Quantization}
    \label{fig:noarb_quantization}
\end{figure}

\begin{figure}

    \begin{subfigure}{0.9\textwidth}
        \centering
        \includegraphics[width=0.9\linewidth]{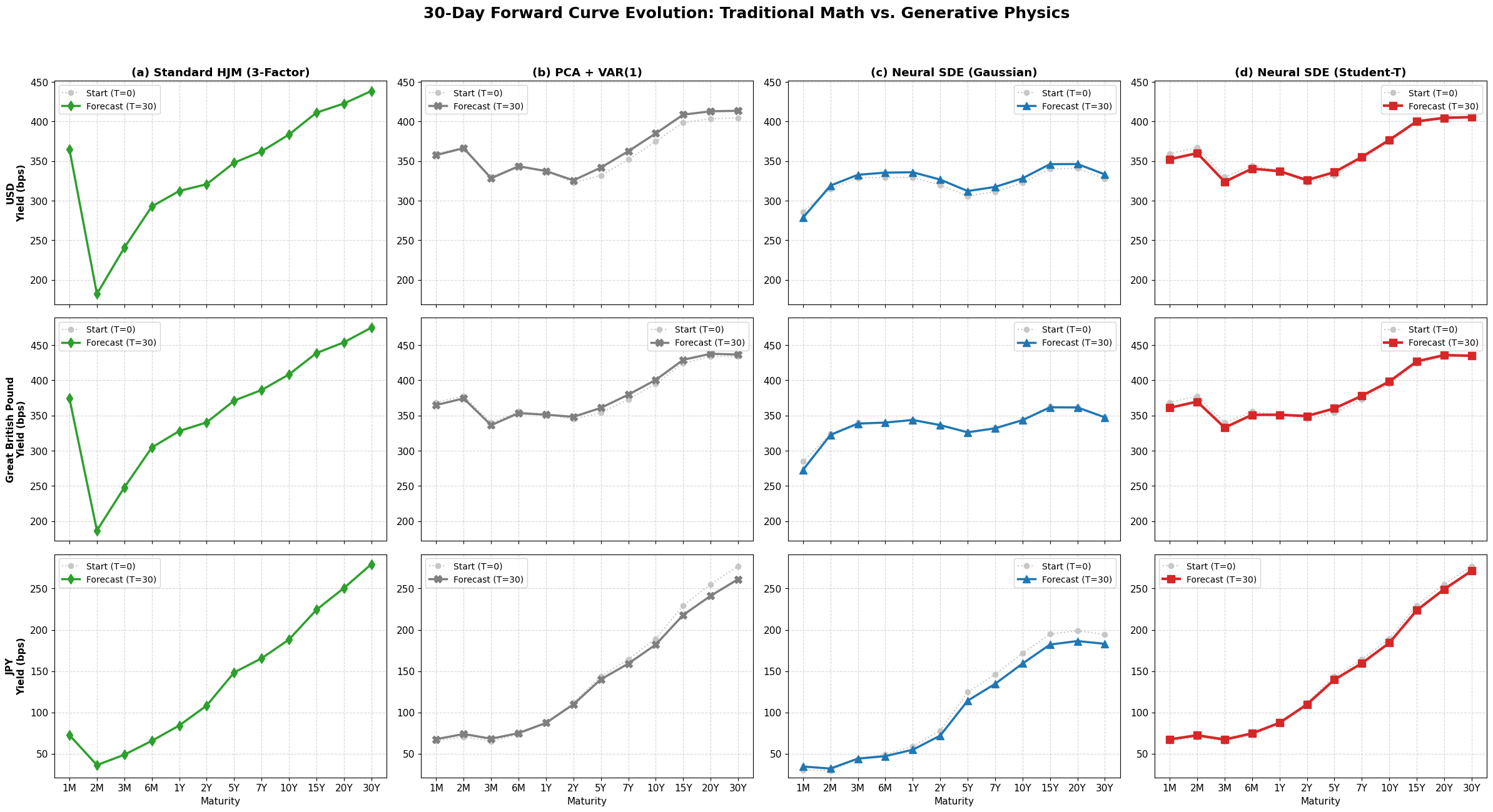}
        \caption{\textbf{30-Day Forward Curve Evolution Stress Test.} Out-of-sample projections across diverse macroeconomic regimes (USD, GBP, JPY). (a) Standard HJM exhibits massive parallel drift and violates Japan's zero-lower-bound. (b) PCA + VAR(1) produces unnatural distortions due to missing physical anchors. (c) Standard Gaussian SDE over-smooths the term structure, indicating manifold collapse. Ultimately, (d) the proposed Student-t Neural SDE uniquely preserves structural resilience. Driven by non-linear physical gravity, it enforces arbitrage-free constraints and local monetary bounds without sacrificing idiosyncratic curve convexity.}
    \end{subfigure}

    \vspace{0.5cm}

    \begin{subfigure}{0.9\textwidth}
        \centering
        \includegraphics[width=0.9\linewidth]{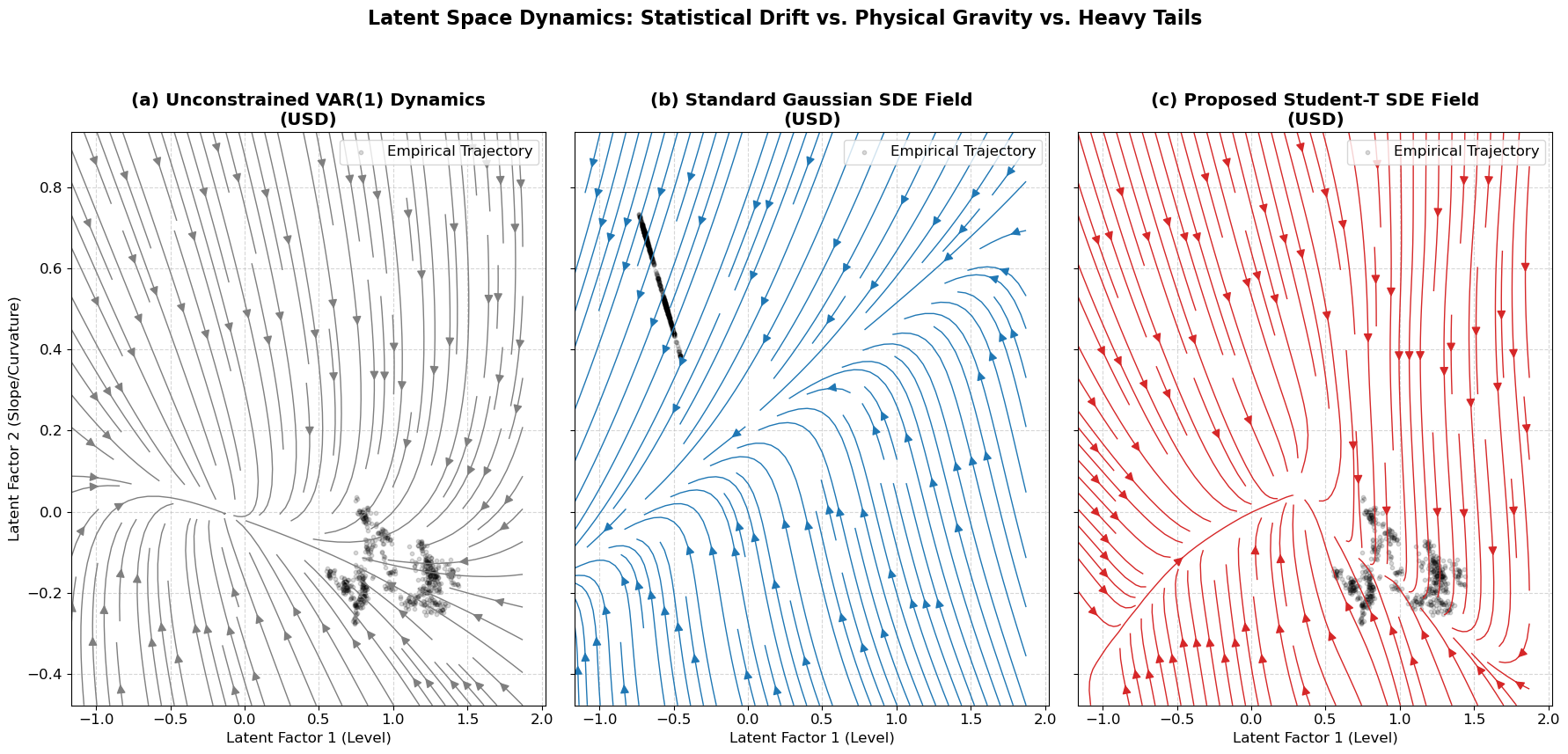}
        \caption{\textbf{Latent Phase Space Dynamics and Trajectory Evolution.} Top row: learned velocity vector fields; Bottom row: simulated latent trajectories. (a) Unconstrained VAR(1) exhibits chaotic drift and unbounded statistical noise. (b) Standard Gaussian SDE suffers from manifold collapse, forcing trajectories into a rigid 1D diagonal and over-smoothing outputs. (c) The proposed Student-t Neural SDE successfully preserves a 3D manifold topology. It applies a non-linear physical gravity that strictly bounds trajectories to arbitrage-free states without crushing essential idiosyncratic flexibility.}
    \end{subfigure}

    \caption{\textbf{Comprehensive Stage B Evaluation}}
    \label{fig:three_vertical_plots}
\end{figure}

\textbf{Evaluating Stage B Dynamics:} To validate the continuous-time modeling, we subject our proposed Neural SDE to a comparative analysis (Figure \ref{fig:noarb_quantization} and \ref{fig:three_vertical_plots}) against a Standard 3-Factor HJM, an unconstrained PCA+VAR, and a Standard Gaussian Neural SDE.
\begin{enumerate}
    \item \textit{The Breakdown of Traditional HJM and VAR:} The Standard HJM model completely fails to maintain macroeconomic realism out-of-sample, suffering from massive parallel downward drifts in aggressive rate environments. Bound by Gaussian Brownian motion, it completely violates the Yield Curve Control (YCC) regime in Japan, projecting JPY yields to spike past 150 bps, ignoring the physical reality of the zero-lower-bound. The unconstrained VAR(1) baseline similarly suffers from severe ``term structure drift,'' projecting unnatural kinks.
    \item \textit{The Flaw of Gaussian Over-Smoothing:} The Standard Gaussian SDE eliminates kinks by enforcing PDE smoothness but introduces manifold collapse. The phase space vector fields reveal that the penalty is overly rigid, crushing the latent representation into a strict 1D diagonal. It over-smooths curves, washing out idiosyncratic shapes (e.g., the USD hump), and fails to capture extreme convexity adjustments, leading to massive, exponentially growing arbitrage residuals from 5Y to 30Y tenors.
    \item \textit{The Triumph of the Student-t Neural SDE:} The proposed model elegantly resolves this conflict. By decoupling the base rate via LevelScript, it eradicates parallel drifts. By accommodating heavy-tailed shocks, it prevents manifold collapse. The deterministic drift ($\mu_{\mathbb{P}}$) acts as a non-linear "physical gravity." Rather than blindly pulling curves to a historical mean, it smoothly guides extreme curve shapes back to theoretically sound, arbitrage-free states while perfectly respecting local monetary regimes (e.g., remaining bounded under JPY yield curve control).
\end{enumerate}

\begin{figure}[t]
\centering
\includegraphics[width=0.8\textwidth]{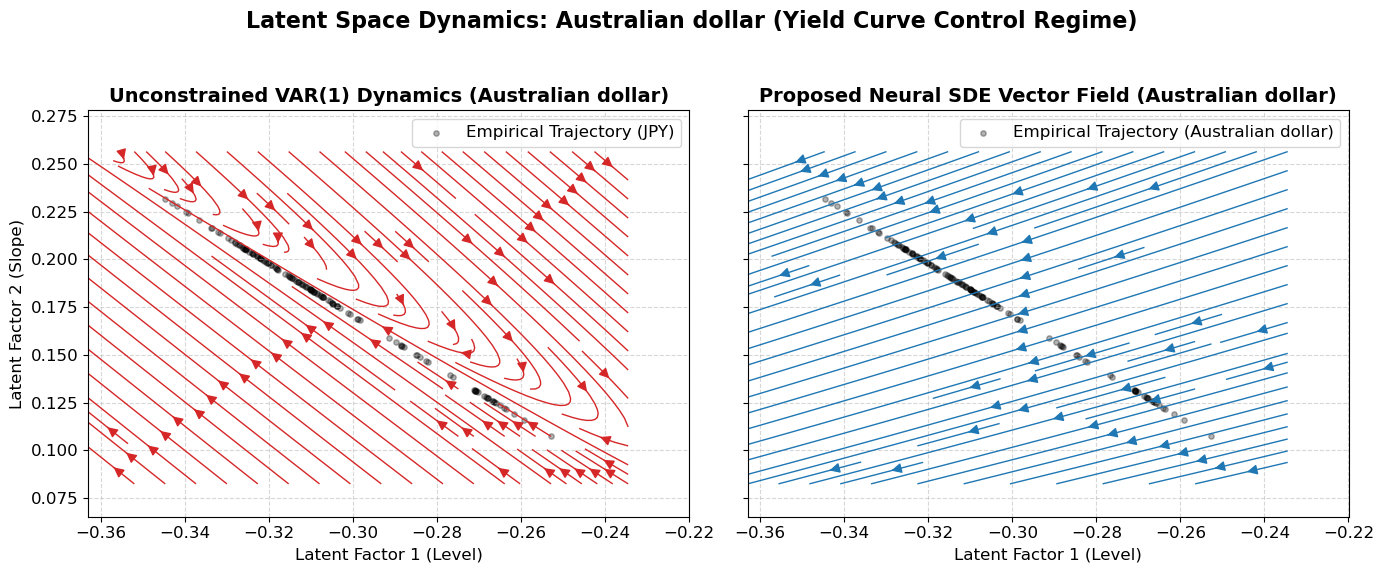}
\caption{Latent space vector fields for the Australian Dollar (AUD) yield curve. Similar to the USD regime, the unconstrained baseline (left) generates chaotic extrapolation trajectories. The Neural SDE (right) maintains strict topological stability under the PDE constraint. This demonstrates that the model's structural regularization is robust across different sovereign bond markets and is not merely overfitted to the liquidity characteristics of the US Treasury market.}
\label{fig:aud_varsde_latentspace}
\end{figure}

\begin{figure}[t]
\centering
\includegraphics[width=0.8\textwidth]{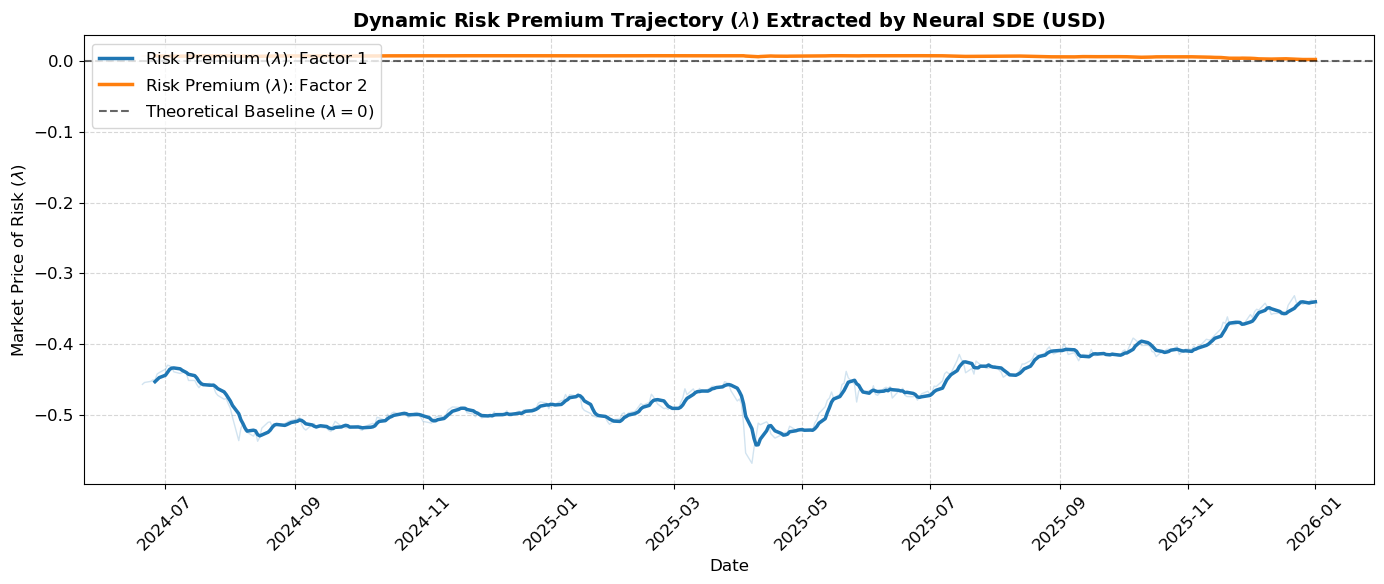}
\caption{Time-varying market price of risk ($\lambda$) extracted by the Neural SDE for the USD market. While the slope factor ($\lambda_2$) remains heavily anchored near the theoretical risk-neutral baseline (zero), the level factor ($\lambda_1$) exhibits significant volatility in the negative domain. This high-frequency oscillation captures the dynamic term premium demanded by investors during the 2024–2025 macroeconomic regime, characterized by intense market speculation regarding the Federal Reserve's monetary policy pivots.}
\label{fig:usd_riskpremium_tsplot}
\end{figure}

\begin{figure}[t]
\centering
\includegraphics[width=0.8\textwidth]{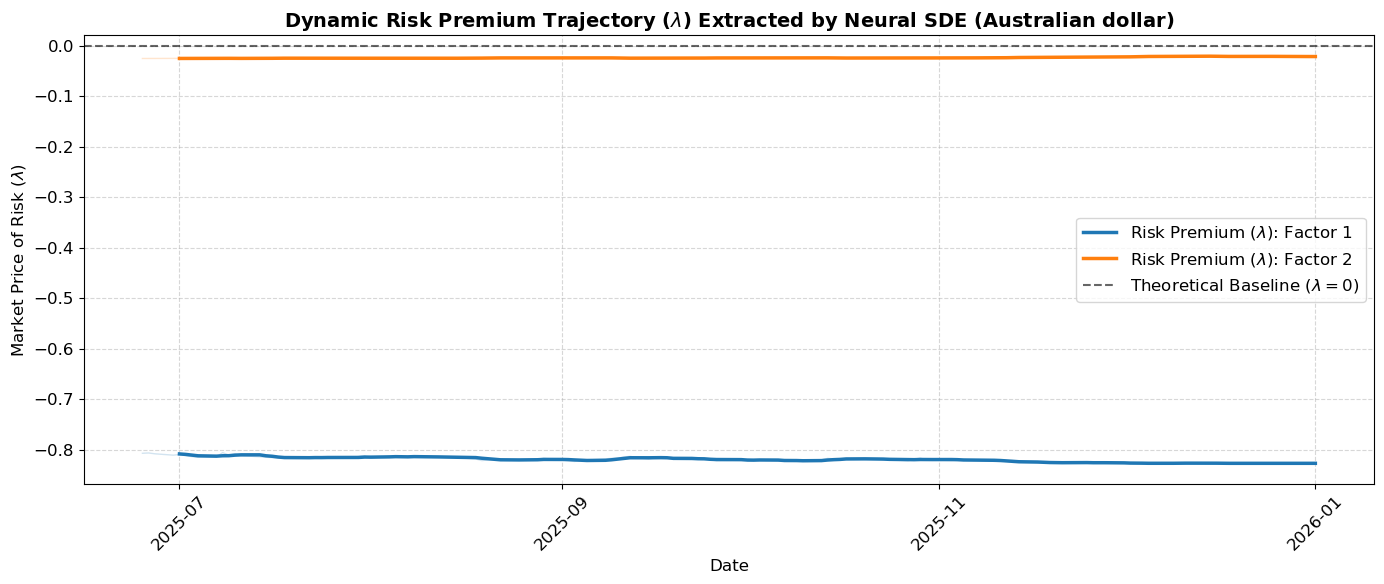}
\caption{Time-varying market price of risk ($\lambda$) extracted by the Neural SDE for the AUD market. In stark contrast to the volatile USD risk premium, the AUD level factor ($\lambda_1$) exhibits a deep, persistent, and highly stable negative discount (approximately $-0.8$). This structural flatness accurately reflects Australia's unique "sticky inflation" macroeconomic regime during this period, where the Reserve Bank of Australia (RBA) maintained prolonged high interest rates, resulting in a stagnant but deeply discounted term premium.}
\label{fig:aud_riskpremium_tsplot}
\end{figure}

\textbf{Macroeconomic Regime Extraction:} Beyond structural stability, Figures \ref{fig:aud_varsde_latentspace}, \ref{fig:usd_riskpremium_tsplot}, and \ref{fig:aud_riskpremium_tsplot} validate the Neural SDE's capacity as an unsupervised macroeconomic regime detector. By isolating the physical drift from the risk-neutral drift, the model extracts the market price of risk ($\lambda$) for each factor. In both economies, the slope/curvature risk premium ($\lambda_2$) tightly hugs the zero-line, aligning with financial intuition that relative shape distortions are rapidly arbitraged away. The diagnostic power lies in the level factor risk premium ($\lambda_1$). For the USD market (Figure \ref{fig:usd_riskpremium_tsplot}), $\lambda_1$ is highly volatile, perfectly aligning with the 2024-2025 macroeconomic reality where bond markets were hyper-sensitive to Fed policy pivots. In stark contrast, the AUD market (Figure \ref{fig:aud_riskpremium_tsplot}) presents a deeply negative (-0.8) and stagnant $\lambda_1$ trajectory. During this window, Australia suffered from persistent "sticky" inflation, forcing the RBA to hold rates at a restrictive plateau. The Neural SDE blindly ingests the data and correctly outputs a "stagnant" risk premium—investors demanded a massive, unchanging compensation for holding Australian debt in an immobilized, high-inflation regime.

%===============================================
%===========Conclusion=======
%===============================================
\section{Conclusion}
\label{sec:conclusion}
In this study, we have introduced a novel physics-informed generative framework that resolves the long-standing dichotomy between the statistical expressivity of deep learning and the rigorous theoretical constraints of fixed-income mathematics. By integrating a two-stage architecture comprising a Student-t Conditional Variational Autoencoder with LevelScript ($\text{CVAE}_{\text{sT}} + \text{LS}$) and a No-Arbitrage Neural SDE, we have demonstrated that it is possible to learn complex, non-linear term structure dynamics without sacrificing the fundamental laws of financial physics. The core methodological contribution of this research lies in the synergy of three structural innovations. First, the introduction of the LevelScript (LS) anchor effectively decouples absolute rate levels from macroeconomic shape dynamics, providing a deterministic "physical anchor" that prevents the catastrophic parallel drifts observed in traditional models. Second, the adoption of a Student-t likelihood within the CVAE architecture ensures that the learned manifold remains robust to heavy-tailed market shocks, thereby preventing "manifold collapse" during periods of extreme volatility. Third, the formulation of a No-Arbitrage PDE penalty within a continuous-time Neural SDE framework ensures that the latent trajectories remain consistent with the risk-neutral pricing measure, providing a data-driven solution to the Filipović consistency problem. Empirical evaluations across diverse sovereign regimes—including USD, GBP, and JPY—confirm the superior performance of the proposed architecture. Our model achieved an exceptional out-of-sample Mean Tenor RMSE of 6.58 bps, significantly outperforming classical 3-factor HJM models and unconstrained statistical baselines (VAR). Most notably, the model demonstrates remarkable structural resilience: it successfully respects the zero-lower-bound in the JPY regime and maintains realistic curve convexity in aggressive hiking cycles, where traditional models often suffer from structural drift or over-smoothing. Beyond forecasting accuracy, this framework establishes a new benchmark for continuous-time scenario generation and unsupervised regime detection. The learned phase-space vector fields reveal a sophisticated "physical gravity" that guides term structures back to theoretically sound states, providing quantitative strategists with a robust engine for stress testing and risk management. In conclusion, our research proves that physics-informed generative models represent the next evolution in interest rate modeling. By replacing rigid parametric assumptions with data-driven manifold learning—while strictly enforcing no-arbitrage principles—this architecture provides a scalable and mathematically rigorous foundation for the next generation of global fixed-income analysis.

\subsection{Limitations and Further Study}

Despite its robust performance across multiple sovereign regimes, our framework presents several theoretical and empirical limitations that offer promising avenues for future research:

\subsubsection{Data Insufficiency}

We utilize Overnight Index Swap rates to construct our model, as these rates are publicly quoted and inherently arbitrage-free due to market liquidity. However, a significant challenge arises because many currencies do not exhibit synchronized exposure timelines, making it difficult to standardize data across different currencies. Our dataset, sourced from Refinitiv, is often limited in the number of data points, requiring us to truncate incomplete datasets using the methodologies outlined in 
section \ref{sec:appendix} to prioritize meaningful information. Additionally, we posit that with access to a more extensive and robust dataset, our model's accuracy would improve, enabling it to better capture global interest rate dynamics across various currencies.

\subsubsection{The Zero-Lower-Bound (ZLB) and Extreme Idiosyncratic Regimes}

As observed in our out-of-sample evaluation, the Japanese Yen (JPY) continues to pose unique challenges. While the proposed model vastly outperforms baselines, the residual errors in JPY indicate that extreme, prolonged Yield Curve Control (YCC) or zero-lower-bound environments strain universal generative models. The symmetrical nature of the Student-t distribution, while excellent for capturing heavy tails, does not explicitly account for the asymmetric boundary condition of near-zero rates. Future studies could explore integrating asymmetric likelihood functions (e.g., log-normal or specialized shadow-rate transformations) within the decoder to natively enforce zero or negative lower bounds.

\subsubsection{Integration of Explicit Macroeconomic Covariates}

Currently, the Neural SDE acts as an unsupervised regime detector, inferring the market price of risk ($\lambda$) purely from the latent evolution of the yield curve itself. While this data-driven approach is elegant, term structure dynamics are fundamentally driven by central bank policies and macroeconomic indicators. Further research could extend the drift networks $\mu_{\mathbb{P}}(z)$ and $\mu_{\mathbb{Q}}(z)$ to incorporate exogenous macroeconomic covariates (e.g., inflation prints, central bank policy rates, or textual sentiment from FOMC meetings). Embedding these structural variables as external forcing terms in the SDE could further constrain long-horizon forecasting and improve the interpretability of the extracted risk premiums.

\subsubsection{Comprehensive Benchmarking against Advanced AEMM Architectures}
While the empirical evaluation in this study demonstrates the superiority of our framework over classical HJM and unconstrained statistical models, the generative finance literature continues to evolve rapidly. In Section \ref{sec:benchmarks}, we formalized two advanced theoretical benchmarks: a DNS-style Autoencoder ($\mathbb{P}$-measure) and a Forward-Rate AEMM utilizing a Decoder-Jacobian Volatility Basis ($\mathbb{Q}$-measure). The latter model, which forces trajectories onto the manifold via discrete re-encoding projections, represents an elegant alternative to our continuous-time Neural SDE penalty approach. A highly valuable direction for further study would be a large-scale, head-to-head empirical replication and comparison between our Student-t constrained SDE and these Jacobian-basis projection models. Such an analysis would provide deeper insights into the trade-offs between continuous PDE penalization and discrete manifold projection under extreme macroeconomic stress.

%===============================================
%===========Figures & Tables=======
%===============================================
\section*{Figures and Tables}

% table 1
\begin{table}[h]
    \centering
    \renewcommand{\arraystretch}{1.5} % Increases row height for readability
    \caption{Summary of Mathematical Linkages Between Interest Rate Measures}
    \begin{tabular}{l c l}
        \toprule
        \textbf{Measure} & \textbf{Notation} & \textbf{Relation to Bond Price} $P(t, T)$ \\
        \midrule
        Zero-Coupon Yield & $y(t, T)$ & $P(t, T) = \exp\left(-y(t, T) \cdot (T - t)\right)$ \\
        Instantaneous Forward Rate & $f(t, T)$ & $P(t, T) = \exp\left(-\int_t^T f(t, u) \, du\right)$ \\
        Par Swap Rate & $S(t)$ & $S(t) = \frac{P(t, T_0) - P(t, T_n)}{\sum_{i=1}^{n} \tau_i P(t, T_i)}$ \\
        \bottomrule
    \end{tabular}
    \label{tab:mathrelation_interestrates}
\end{table}

\begin{table}[htbp] % [htbp] 确保表格紧跟文字，不乱跑到文章末尾
\centering
\caption{Ablation Test: Out-of-Sample RMSE Comparison Across Models (in bps). The proposed synergistic framework ($CVAE_{sT} + LS$) dramatically outperforms both classical benchmarks and standard generative architectures across all tenors and regimes.}
\label{tab:ablation}
\resizebox{\textwidth}{!}{% 自动缩放以适应页面宽度
\begin{tabular}{@{}lcccccccc@{}} % @{} 去除两端多余空白
\toprule
\textbf{Error (RMSE in bps)} & \textbf{VAE$_{\mathcal{N}}$} & \textbf{VAE$_{sT}$} & \textbf{CVAE$_{\mathcal{N}}$} & \textbf{CVAE$_{\mathcal{N}}$ + LS} & \textbf{CVAE$_{sT}$} & \textbf{CVAE$_{sT}$ + LS (Proposed)} & \textbf{PCA} & \textbf{FinQ} \\ \midrule
\multicolumn{9}{c}{\textit{Panel A: RMSE by Maturity Tenor}} \\ \midrule
1 Month  & 112.49 & \textbf{9.13}  & 31.75  & 37.92  & 60.16 & 11.19 & 2842.43 & 181.37 \\
6 Month  & 112.27 & \textbf{4.59}  & 24.31  & 20.42  & 5.50  & 5.90  & 2859.21 & 170.32 \\
1 Year   & 99.00  & 90.50 & 24.58  & 19.33  & 5.56  & \textbf{2.89}  & 2673.55 & 160.64 \\
7 Year   & 69.06  & 49.26 & 26.24  & 19.42  & 3.30  & \textbf{2.95}  & 1452.28 & 135.39 \\
15 Year  & 72.87  & 47.66 & 37.04  & 19.28  & 2.65  & \textbf{2.32}  & 1302.70 & 134.49 \\
30 Year  & 83.53  & 55.00 & 56.56  & 34.26  & 6.65  & \textbf{5.13}  & 1026.67 & 132.74 \\ \midrule
\textbf{Mean (All Tenors)} & 89.51 & 29.04 & 31.30 & 23.45 & 12.02 & \textbf{6.58} & 2007.24 & 152.11 \\ \midrule
\multicolumn{9}{c}{\textit{Panel B: RMSE by Sovereign Currency}} \\ \midrule
USD      & 397.60 & 18.42 & 423.37 & 37.73  & 18.31 & \textbf{7.83}  & 4308.95 & 293.18 \\
JPY      & 123.76 & 43.89 & 74.73  & 18.04  & 7.18  & \textbf{2.42}  & 5230.82 & 44.11  \\
GBP      & 404.60 & 22.05 & 410.67 & 39.71  & 18.00 & \textbf{8.13}  & 1669.49 & 297.45 \\
CHF      & 96.02  & 30.25 & 58.67  & 12.77  & 9.16  & \textbf{3.31}  & 253.84  & 90.12  \\
CAD      & 242.66 & 41.05 & 262.74 & 15.25  & 10.36 & \textbf{5.94}  & 2068.63 & 141.22 \\ \midrule
\textbf{Mean (All Currencies)} & 245.05 & 32.29 & 243.85 & 23.44 & 12.02 & \textbf{5.36} & 2007.24 & 152.11 \\ \bottomrule
\end{tabular}
}
\end{table}

% 1

% Parameters Table
\begin{table}[H]
    \centering
    \caption{Detailed Neural Network Architecture}
    \label{tab:architecture}
    \small
    \begin{tabular}{@{}lllc@{}}
    \toprule
    \textbf{Module} & \textbf{Layer Type} & \textbf{Dimensions (In $\rightarrow$ Out)} & \textbf{Activation} \\ \midrule
    \multicolumn{4}{c}{\textbf{Stage A: CVAE$_{\text{sT}}$ Encoder $q_\phi(z | x, c, l)$}} \\ \midrule
    Input Concat    & -                   & $12 \text{ (Shape)} + 1 \text{ (Level)} + 8 \text{ (Curr)}$ & -                   \\
    Hidden Block 1  & Linear + BatchNorm  & $21 \rightarrow 128$                       & Softplus            \\
    Hidden Block 2  & Linear + BatchNorm  & $128 \rightarrow 64$                       & Softplus            \\
    Output Head $\mu_z$ & Linear              & $64 \rightarrow 3$                         & Linear              \\
    Output Head $\log\sigma_z^2$ & Linear     & $64 \rightarrow 3$                         & Linear              \\ \midrule
    \multicolumn{4}{c}{\textbf{Stage A: Student-t Decoder $p_\theta(x | z, c)$}} \\ \midrule
    Input Concat    & -                   & $3 \text{ (Latent } z) + 8 \text{ (Curr)}$ & -                   \\
    Hidden Block 1  & Linear              & $11 \rightarrow 64$                        & Softplus            \\
    Hidden Block 2  & Linear              & $64 \rightarrow 128$                       & Softplus            \\
    Output Head $\mu_\theta$ & Linear     & $128 \rightarrow 12$                       & Linear              \\
    Output Head $\lambda_\theta$ (Scale) & Linear + Softplus & $128 \rightarrow 12$        & Softplus ($>0$)     \\
    Output Head $\nu_\theta$ (DoF) & Linear + Softplus & $128 \rightarrow 1$ (Shared) or 12 & Softplus ($>2$)     \\ \midrule
    \multicolumn{4}{c}{\textbf{Stage B: Neural SDE ParamNet}} \\ \midrule
    Input           & -                   & $3 \text{ (Latent } z) + 8 \text{ (Curr)}$ & -                   \\
    Hidden Block 1  & Linear              & $11 \rightarrow 128$                       & Tanh                \\
    Hidden Block 2  & Linear              & $128 \rightarrow 128$                      & Tanh                \\
    Drift $\mu_{\mathbb{P}}$ & Linear     & $128 \rightarrow 3$                        & Linear              \\
    Volatility $\Sigma$ & Linear + Softplus & $128 \rightarrow 3 \times 3$ (Diagonal)  & Softplus ($>0$)     \\
    Risk Premium $\lambda$ & Linear       & $128 \rightarrow 3$                        & Linear              \\ \bottomrule
    \end{tabular}
\end{table}

% 2

% ==========================================
% STAGE A: CVAEsT + LevelScript (上半场)
% ==========================================

%3

%4

%5

% In-Sample

% Out-Of-Sample

%10

%12

%13

%14

%17

%18

%18

\clearpage
%========================================

\section*{Appendix} \label{sec:appendix}

\subsection*{Methods for handling null values in the multi-currency dataset} 
\paragraph{Stable-coverage truncation via rolling completeness}

For each currency $c$, let $r_t^{(c)} \in \mathcal{R}^{\lvert \tau  \rvert}$ denote the tenor vector on date $t$, with potentially missing components. We define the row completeness ratio
\[
\rho_t^{(c)}=\frac{1}{|\mathcal T|}
\sum_{\tau\in\mathcal T}
\mathbf{1}_{\{\, r_t^{(c)}(\tau)\ \text{is observed}\,\}}
\]
We then determine a currency-specific stable start date $T_{start}^{(c)}$ by requiring that completeness remains consistently high over a forward rolling window of length $W$ trading days. Concretely, for thresholds $\rho_0 \in (0, 1)$ and $\pi_0 \in (0, 1)$, we find the earliest date such that within the subsequent window,
\[
\frac{1}{W}\sum_{s=t}^{t+W-1}\mathbf{1}\left\{\rho^{(c)}_s \ge \rho_0\right\}\ge \pi_0 .
\]

In our implementation, this procedure is chosen to eliminate spurious “starts” caused by isolated bursts of availability; it ensures that the retained period reflects a regime where the tenor grid is structurally supported. Finally, each currency is truncated to
\[
t \geq T_{start}^{(c)}
\]
This step directly addresses the dominant source of sparsity—missing values due to early-history coverage—without imposing artificial imputation.

\paragraph{Densification and final quality screening}
After truncation, we retain only dates whose tenor vector is fully observed on 
$T$, i.e., 
\[
r_t^{(c)} \text{ is kept} \iff \rho_t^{(c)} = 1
\]
Equivalently, we apply a strict row-wise filter on the selected tenor columns to drop null values. Because the stable-coverage truncation is performed first, this densification step discards only a small fraction of post-start observations while producing a training table with no missing values.

\subsection*{Theoretical Foundation: Consistency with the Filipović HJM Manifold} \label{sec:appendix_theory}
The empirical success of our two-stage architecture can be mathematically grounded in the consistency theory for Heath-Jarrow-Morton (HJM) models, originally formalized by Filipović (2001) \cite{filipovic2001consistency}. Filipović investigated the conditions under which the infinite-dimensional stochastic partial differential equation (SPDE) shaping the forward rate curve (the Musiela parameterization) admits a finite-dimensional realization.
\paragraph{The Analytical Consistency Problem}
Let $H_w$ be a suitably chosen Hilbert space of forward curves (the Filipović space). The evolution of the forward curve $r_t(x)$ in time-to-maturity coordinates $x = T-t$ is given by the Musiela SPDE:$$dr_t(x) = \left( \frac{\partial}{\partial x}r_t(x) + \sum_{i=1}^n \sigma_i(t,x)\int_0^x \sigma_i(t,u)du \right)dt + \sum_{i=1}^n \sigma_i(t,x)dW_{i,t}^{\mathbb{Q}}$$Filipović proved that a finite-dimensional parameterized manifold $\mathcal{M} = \{ G(z) \mid z \in \mathcal{Z} \subset \mathbb{R}^d \}$ is invariant (or consistent) with the HJM dynamics if and only if the drift and volatility vector fields of the SPDE are strictly tangent to $\mathcal{M}$ at every point. Historically, this condition was overly restrictive for analytical parametric forms. For instance, Filipović famously proved that the widely used Nelson-Siegel (NS) family does not form a consistent manifold, meaning that a NS curve will almost surely evolve into a non-NS curve under arbitrage-free dynamics.
\paragraph{The Neural Manifold as a Data-Driven Solution} 
Our Stage A architecture (CVAEsT + LS) provides a non-linear, data-driven solution to this exact problem. Rather than guessing an analytical family $G(z)$ like exponential-polynomials, we use the frozen, trained decoder $D^{(P)}(z, \tau)$ to define a highly expressive, $d$-dimensional ($3$ as the best case in our framework) neural manifold:$$\mathcal{M}_{Neural} = \left\{ D^{(P)}(z, \cdot) \mid z \in \mathbb{R}^d \right\}$$Because this manifold is learned from the empirical OIS swap data, it inherently spans the true topological space of observed macroeconomic curve shapes, avoiding the rigidity of analytical parameterizations. 
\paragraph{Stage B as the Filipović Tangency Penalty}
In Stage B, we drive the latent state $z_t$ using a Neural SDE. For $\mathcal{M}_{Neural}$ to be a consistent HJM manifold under Filipović's definition, the stochastic evolution of the discounted bond prices generated by $D^{(P)}$ must be a local martingale under the risk-neutral measure $\mathbb{Q}$.Our No-Arbitrage PDE loss ($\mathcal{L}_{arb}$) is the direct empirical translation of Filipović’s tangency requirement. Recall the PDE residual defined in Section \ref{sec:lossfunc}:$$\mathcal{L}_{arb} = -\partial_\tau D^{(P)} + \nabla_z D^{(P)\top} \mu_{\mathbb{Q}} + \frac{1}{2}\text{Tr}\left(\Sigma\Sigma^\top \nabla_{zz}^2 D^{(P)}\right) - r_t D^{(P)} = 0$$By penalizing this residual, the network continuously adjusts the risk-neutral drift $\mu_{\mathbb{Q}}$ and volatility $\Sigma$ such that the resulting local dynamics of the bond price lie perfectly within the tangent bundle spanned by the decoder's Jacobian $\nabla_z D^{(P)}$. Therefore, our framework is not merely applying a smoothing penalty; it is computationally solving Filipović's consistency problem. While traditional parametric models fail the consistency test (forcing practitioners to introduce continuous recalibration), our model learns a finite-dimensional neural manifold that is both empirically accurate (Stage A) and theoretically invariant under continuous-time no-arbitrage dynamics (Stage B).

\section*{Acknowledgments}
\bibliographystyle{elsarticle-num} % 这是 Elsevier 官方的纯数字格式

% 第二行：指定你的文献数据库文件名
\bibliography{references} % ⚠️ 注意：这里只写名字，千万不要加 .bib 后缀！

@misc{akiba2019optunanextgenerationhyperparameteroptimization,
      title={Optuna: A Next-generation Hyperparameter Optimization Framework}, 
      author={Takuya Akiba and Shotaro Sano and Toshihiko Yanase and Takeru Ohta and Masanori Koyama},
      year={2019},
      eprint={1907.10902},
      archivePrefix={arXiv},
      primaryClass={cs.LG},
      url={https://arxiv.org/abs/1907.10902}, 
}

@article{hutchinson1989stochastic,
  title={A stochastic estimator of the trace of the influence matrix for Laplacian smoothing splines},
  author={Hutchinson, Michael F},
  journal={Communications in Statistics-Simulation and Computation},
  volume={18},
  number={3},
  pages={1059--1076},
  year={1989},
  publisher={Taylor \& Francis}
}

@article{pearlmutter1994fast,
  title={Fast exact multiplication by the Hessian},
  author={Pearlmutter, Barak A},
  journal={Neural computation},
  volume={6},
  number={1},
  pages={147--160},
  year={1994},
  publisher={MIT Press}
}

@article{diebold2006forecasting,
  title={Forecasting the term structure of government bond yields},
  author={Diebold, Francis X and Li, Canlin},
  journal={Journal of econometrics},
  volume={130},
  number={2},
  pages={337--364},
  year={2006},
  publisher={Elsevier}
}

@book{filipovic2001consistency,
  title={Consistency problems for Heath-Jarrow-Morton interest rate models},
  author={Filipovic, Damir},
  year={2001},
  publisher={Springer Science \& Business Media}
}

@article{davidson2018hyperspherical,
  title={Hyperspherical variational auto-encoders},
  author={Davidson, Tim R and Falorsi, Luca and De Cao, Nicola and Kipf, Thomas and Tomczak, Jakub M},
  journal={arXiv preprint arXiv:1804.00891},
  year={2018}
}

@inproceedings{takahashi2018student,
  title={Student-t Variational Autoencoder for Robust Density Estimation.},
  author={Takahashi, Hiroshi and Iwata, Tomoharu and Yamanaka, Yuki and Yamada, Masanori and Yagi, Satoshi},
  booktitle={IJCAI},
  pages={2696--2702},
  year={2018}
}

@inproceedings{boier2023multiresolution,
  title={Multiresolution Signal Processing of Financial Market Objects},
  author={Boier, Ioana},
  booktitle={ICASSP 2023-2023 IEEE International Conference on Acoustics, Speech and Signal Processing (ICASSP)},
  pages={1--5},
  year={2023},
  organization={IEEE}
}

@article{nelson1987parsimonious,
  title={Parsimonious modeling of yield curves},
  author={Nelson, Charles R and Siegel, Andrew F},
  journal={Journal of business},
  pages={473--489},
  year={1987},
  publisher={JSTOR}
}

@inproceedings{saha2025matching,
  title={Matching aggregate posteriors in the variational autoencoder},
  author={Saha, Surojit and Joshi, Sarang and Whitaker, Ross},
  booktitle={International Conference on Pattern Recognition},
  pages={428--444},
  year={2025},
  organization={Springer}
}

@article{dutta2025learning,
  title={Learning Energy-based Variational Latent Prior for VAEs},
  author={Dutta, Debottam and Amballa, Chaitanya and Xu, Zhongweiyang and Wei, Yu-Lin and Choudhury, Romit Roy},
  journal={arXiv preprint arXiv:2510.00260},
  year={2025}
}

@article{kingma2013auto,
  title={Auto-encoding variational bayes},
  author={Kingma, Diederik P and Welling, Max},
  journal={arXiv preprint arXiv:1312.6114},
  year={2013}
}

@article{dybvig1996long,
  title={Long forward and zero-coupon rates can never fall},
  author={Dybvig, Philip H and Ingersoll Jr, Jonathan E and Ross, Stephen A},
  journal={Journal of Business},
  pages={1--25},
  year={1996},
  publisher={JSTOR}
}

@article{sokol2022autoencoder,
  title={Autoencoder market models for interest rates},
  author={Sokol, Alexander},
  journal={Available at SSRN 4300756},
  year={2022}
}

@article{litterman1991volatility,
  title={Volatility and the yield curve},
  author={Litterman, Robert B and Scheinkman, Jos{\'e} and Weiss, Laurence},
  journal={The Journal of Fixed Income},
  volume={1},
  number={1},
  pages={49--53},
  year={1991},
  publisher={Pageant Media US}
}

@article{christensen2011affine,
  title={The affine arbitrage-free class of Nelson--Siegel term structure models},
  author={Christensen, Jens HE and Diebold, Francis X and Rudebusch, Glenn D},
  journal={Journal of Econometrics},
  volume={164},
  number={1},
  pages={4--20},
  year={2011},
  publisher={Elsevier}
}

@article{Andreasen2023Decoding,
  author  = {Andreasen, Jesper},
  title   = {Decoding the Autoencoder},
  journal = {Wilmott},
  year    = {2023},
  volume  = {2023},
  number  = {127},
  doi     = {10.54946/wilm.11166},
  url     = {https://doi.org/10.54946/wilm.11166}
}

@misc{bergeron2021variationalautoencodershandsoffapproach,
      title={Variational Autoencoders: A Hands-Off Approach to Volatility}, 
      author={Maxime Bergeron and Nicholas Fung and John Hull and Zissis Poulos},
      year={2021},
      eprint={2102.03945},
      archivePrefix={arXiv},
      primaryClass={q-fin.CP},
      url={https://arxiv.org/abs/2102.03945}, 
}

@article{lyashenko2024autoencoder,
  title={Autoencoder-Based Risk-Neutral Model for Interest Rates},
  author={Lyashenko, Andrei and Mercurio, Fabio and Sokol, Alexander},
  journal={Available at SSRN 4836728},
  year={2024}
}

@misc{svensson1994estimating,
  title={Estimating and interpreting forward interest rates: Sweden 1992-1994},
  author={Svensson, Lars EO},
  year={1994},
  publisher={National bureau of economic research Cambridge, Mass., USA}
}

@article{heath1992bond,
  title={Bond pricing and the term structure of interest rates: A new methodology for contingent claims valuation},
  author={Heath, David and Jarrow, Robert and Morton, Andrew},
  journal={Econometrica: Journal of the Econometric Society},
  pages={77--105},
  year={1992},
  publisher={JSTOR}
}

@book{musiela1993different,
  title={Different dynamical specifications of the term structure of interest rates and their implications},
  author={Musiela, Marek and Sondermann, Dieter and others},
  year={1993},
  publisher={Rheinische Friedrich-Wilhelms-Universit{\"a}t Bonn}
}

@article{vasicek1977equilibrium,
  title={An equilibrium characterization of the term structure},
  author={Vasicek, Oldrich},
  journal={Journal of financial economics},
  volume={5},
  number={2},
  pages={177--188},
  year={1977},
  publisher={Elsevier}
}

@article{cox1985theory,
  title={A theory of the term structure of interest rates},
  author={Cox, John C and Ingersoll, Jonathan E and Ross, Stephen A and others},
  journal={Econometrica},
  volume={53},
  number={2},
  pages={385--407},
  year={1985},
  publisher={World Scientific}
}

\end{document}